\documentclass[a4paper]{article}

\usepackage{amsmath,amsfonts,amssymb}
\usepackage{amsthm}
\usepackage{enumerate}
\usepackage{graphicx}
\usepackage{rotating}
\usepackage{hyperref}
\usepackage{cite}

\def\ds{\displaystyle}
\def\bea{\begin{array}{c}}
\def\ea{\end{array}}
\def\be{\begin{equation}\bea\ds}
\def\ee{\ea\end{equation}}
\def\bee{\begin{equation}\begin{array}{rcl}\ds}
\def\eee{\end{array}\end{equation}}

\author{Arthur G.~Cavalcanti$^{a}$ and Dmitry Melnikov$^{b,c}$}
\title{Bubble quenches in the AdS/BCFT model}

\begin{document}

\maketitle

\vspace{-5cm}
\hfill{ITEP-TH-10/20}
\vspace{5cm}

\vspace{-30pt}
\begin{center}
\textit{\small $^a$  Department of Theoretical and Experimental Physics, Federal University of Rio Grande do Norte, \\ Campus Universit\'ario, Lagoa Nova, Natal-RN  59078-970, Brazil}
\\ \vspace{6pt}
\textit{\small $^b$  International Institute of Physics, Federal University of Rio Grande do Norte, \\ Campus Universit\'ario, Lagoa Nova, Natal-RN  59078-970, Brazil}\\ \vspace{6pt}
\textit{\small $^c$  Institute for Theoretical and Experimental Physics, \\B.~Cheremushkinskaya 25, Moscow 117218, Russia}
\\ \vspace{2cm}
\end{center}

\vspace{-2cm}

\begin{abstract}
In this paper we construct time-dependent solutions of three-dimensional gravity in AdS space dual to systems with boundaries (BCFTs), following the AdS/BCFT prescription. Such solutions can be discussed in the context of the dynamics of first order phase transitions, or more generally, in the description of quantum quenches. As an example, we apply the holographic model to calculate the dynamics of the entanglement entropy of a local quench corresponding to a nucleation of a Euclidean bubble. As in the known 1+1 CFT examples of local cut and glue quenches, the holographic entanglement entropy grows logarithmically with time with the correct universal coefficient. However, in the bubble quench, the behavior is different at late times. The AdS/BCFT model exhibits the light-cone spreading of correlations and saturation at late times. We also find an analytical formula for the entropy at finite temperature. In the latter case the initial logarithmic growth is followed by the linear law at intermediate times.
\end{abstract}

\maketitle
\vspace{50pt}

\section{Introduction}

Conformal field theories (CFT)~\cite{di1996conformal} often serve as laboratories for interesting phenomena in physics. High degree of symmetry imposes strong constraints not only on the observables of a given theory, but also on its self-consistency. Such constraints allow better analytic control and qualitative understanding of complex phenomena, notably in the strong coupling regime. It is sometimes also possible to generalize the CFT results and techniques beyond the conformal case.

One particular class of interesting problems successfully tackled by the CFT methods in the last couple of decades is the problems related to out-of-equilibrium dynamics of quantum systems~\cite{polkovnikov2011colloquium,eisert2015quantum,d2016quantum,gogolin2016equilibration,altman2015non,Calabrese:2016xau}. Quite often one is interested in the quantum evolution of otherwise stationary system after a quench -- a non-adiabatic change of the Hamiltonian. Besides the general importance of such problems the recent interest was also stimulated by the progress in experimental techniques, such as the control of dynamics of cold atoms~\cite{greiner2002collapse,trotzky2012probing,cheneau2012light,bloch2012quantum,blatt2012quantum,meinert2013quantum,langen2013local,islam2015measuring}, which allowed to test various theoretical model predictions.

CFT methods are particularly powerful in two dimensions. Therefore 1+1 dimensional systems turned a natural subject of the early studies. For example, it is relatively straightforward to analyze time dependence of correlation functions in the quenched systems. Some interesting features of such systems, initially observed in the CFT models, include light-cone spreading of the correlations~\cite{calabrese2006time,Calabrese:2007rg}, linear growth of the entanglement entropy~\cite{Calabrese:2005in} and quantum revivals in finite systems~\cite{Cardy:2014rqa}.

AdS/CFT correspondence~\cite{AdS/CFT,Gubser:1998bc,Witten:1998qj} provides another powerful tool to analyze complex systems, characterized by strong coupling. In low dimension this correspondence can reproduce some characteristic features of conformal theories, especially in two dimensions. A famous example is the holographic formula of Ryu and Takayanagi~\cite{Ryu:2006bv} that expresses the entanglement entropy of a subsystem in terms of the dual geometry. In 1+1 dimensional theory, the entanglement entropy of an interval of length $l$ has a universal piece~\cite{Cardy:1988tk,Holzhey:1994we,Korepin:2004zz,CC}
\be
\label{RT2D}
S_{\rm E} \ = \ \frac{c}{3}\log\frac{l}{\epsilon} + \ldots\,,
\ee
where $c$ is the CFT central charge, $\epsilon$ is the UV cutoff and dots stand 
for the non-universal constant piece. In AdS/CFT the logarithm is a length of 
the geodesic line in the three-dimensional anti de Sitter space, connecting the 
endpoints of the interval. AdS/CFT was also successfully applied to the 
discussion of the quenched dynamics. 
See~\cite{Danielsson:1999fa,Janik:2006gp,AbajoArrastia:2010yt,Albash:2010mv,
Balasubramanian:2010ce,Aparicio:2011zy,Keranen:2011xs,Allais:2011ys,
Buchel:2012gw,Buchel:2013gba,Hartman:2013qma,Liu:2013qca,Bhaseen:2013ypa,
Abajo-Arrastia:2014fma,daSilva:2014zva,Caputa:2013eka} for an 
incomplete set of references. Behavior of correlators and characteristic 
features of the entanglement evolution were reproduced in those studies.

In this work we will be focusing on the discussion of time-dependent dynamics from the point of view of a specific setup in the AdS/CFT correspondence introduced by Takayanagi~\cite{Takayanagi:2011zk}. This setup was dubbed AdS/BCFT as it refers to a gravity dual description of systems with boundaries, which are also amenable to treatment by means of boundary conformal field theories, or BCFTs~\cite{Cardy:1989ir,Cardy:2004hm}. AdS/BCFT exhibits some characteristic features of BCFTs~\cite{Takayanagi:2011zk,AdS/BCFT2,Cavalcanti:2018pta}, although the precise correspondence between the dynamical elements of two approaches has not been established in general.

In AdS/BCFT the dynamics of the boundary of the CFT is encoded in the dynamics 
of codimension one hypersurface ending on that boundary. In what follows we will 
describe new solutions of the AdS/BCFT, which are both time and temperature 
dependent and propose some applications. In particular, we will demonstrate how 
these solutions can be discussed in the context of the evolution of the 
entanglement entropy after a local quench. The quench protocol we will consider 
is similar to the so-called cut and glue quenches. Holographic models of such 
quenches were perhaps originally considered 
in~\cite{Ugajin:2013xxa,Nozaki:2013wia} and more recently 
in~\cite{Mandal:2016cdw, 
Shimaji:2018czt,Caputa:2013eka,Ageev:2019fjf,Kudler-Flam:2020url}. 
Holographic models reproduce well the behavior of the entanglement entropy 
observed in the CFT 
calculations~\cite{Calabrese:2007mtj,2008JSMTE..01..023E,Stephan:2011, 
Stephan:2013,Asplund:2013zba,Asplund:2014coa}. In particular, the entanglement 
entropy grows at initial times, but decays at later ones.

The quench that we will consider exhibits a different entropy behavior. This is related to the fact the initial state is prepared differently. It corresponds to a nucleation of a Euclidean bubble at $t=0$. For $t>0$ the bubble expands. For this reason we will refer to such a protocol as to a bubble quench. 

For early times ${\epsilon}\ll t \ll  l$ the behaviour of the entropy is consistent with the cut and glue quench analysis, as in~\cite{Calabrese:2007mtj}: 
\be
S_{\rm E}(t) \ \sim \ \frac{c}{3}\log\frac{t^2}{\epsilon} + k\,, 
\ee
where $k$ is a non-universal part equal to $k=-(c/3)\log(l/2)$ in our model. At finite temperature the early time behavior is replaced by
\be
S_{\rm E}(t) \ \sim \ \frac{c}{3}\log \left[ \frac{%
2\pi T}{\epsilon }\frac{t^{2}}{\sinh \left(\pi Tl\right)} \right] \,,
\ee
while at intermediate times, $T^{-1}\ll t < l$, it exhibits linear behavior,
\be
S_E\sim \frac{c}{3}2\pi T\left(t-\frac{l}{2}\right) + \frac{c}{3}\log\frac{1}{\pi\epsilon T}\,,
\ee
usually observed in the case of global quenches.

From the standard holographic considerations we argue that at late times $t>\ell$ the entropy should saturate. This occurs in a non-analytic manner (phase transition) in the present model. Such effect is also well known in the analysis of global quenches~\cite{Calabrese:2005in}. It is related to the finite speed of the quasiparticle propagation (light-cone spreading).

This paper is organized as follows. In section~\ref{sec:model} we give a very brief introduction to the AdS/BCFT construction. In section~\ref{sec:solutions} we apply a diffeormphism to construct time-dependent solutions of AdS/BCFT. In section~\ref{sec:examples}
we discuss applications of the solutions to the description of quantum quenches and derive formulae for the time-dependent entanglement entropy and discuss some effects of finite temperature. We summarize our results and observations in section~\ref{sec:conclusions}.

\paragraph*{Note added in the second version.} After the first version of the paper came out we learned about important earlier work on cut and glue quenches, both in the CFT and holographic context. Since our protocol is different from those conventionally used in the cut and glue protocols, we changed the title of the paper and referred to the type of quench studied in this paper as to a bubble quench.

\section{The model}
\label{sec:model}

In this section we are going to briefly define the AdS/BCFT construction of Takayanagi~\cite{Takayanagi:2011zk}. See~\cite{AdS/BCFT2} for a more complete review.

The idea of the construction is to provide a holographic dual description of a system that has a boundary. The boundary can be introduced through boundary conditions imposed on the degrees of freedom. The dual holographic theory must encode the degrees of freedom of the CFT with a boundary, so one does not need the whole of anti de Sitter space to encode the smaller system. Instead one extends the boundary of the CFT into the bulk and introduces additional boundary conditions in the AdS bulk region, which should be compatible with the boundary conditions imposed on the original CFT.

\begin{figure}[tbh]
\centering
\includegraphics[height=6cm]{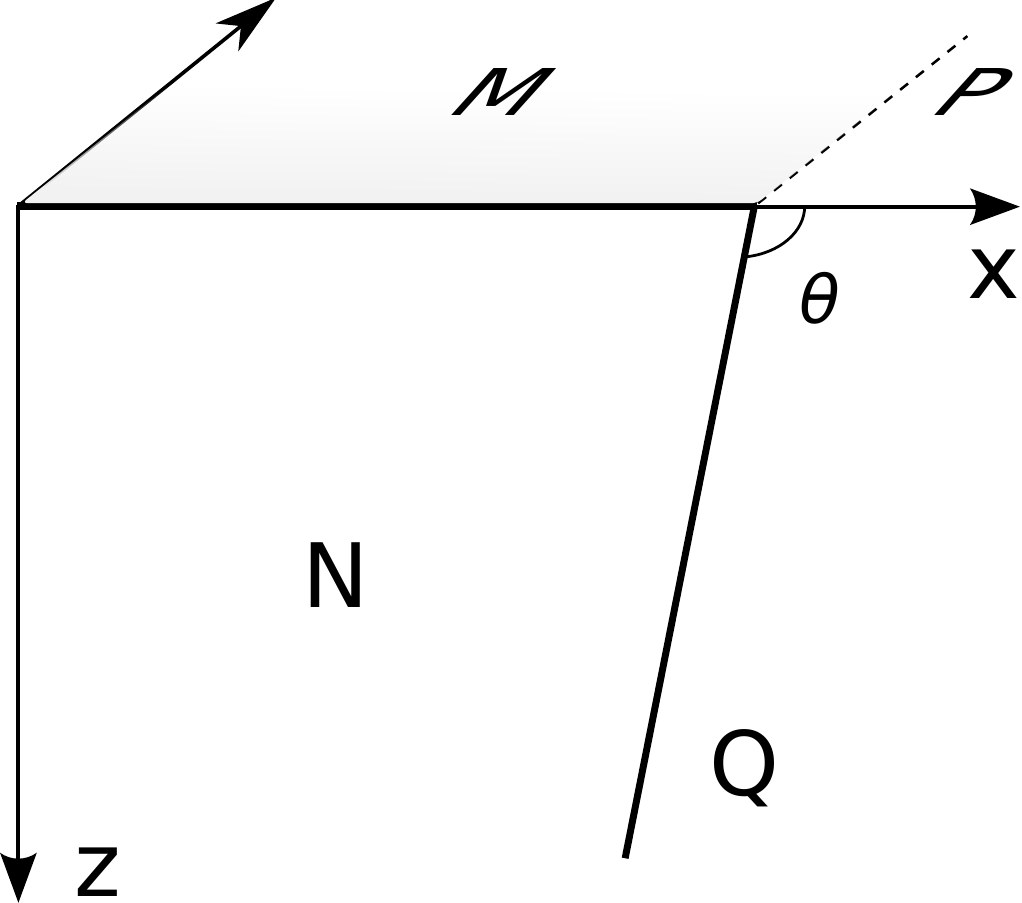} \quad \quad  
\caption{In AdS/BCFT construction of~\cite{Takayanagi:2011zk} gravity theory in $d + 1$-dimensional space $N$ with boundary $Q$ is expected to be dual to a CFT defined in $d$-dimensional space $M$, with boundary $P=\partial M=\partial Q$.}
\label{AdSBCFT}
\end{figure}

The setup is illustrated by figure~\ref{AdSBCFT}. Let the CFT be defined in space $M$ of some dimension $d$, whose boundary is $\partial M=P$. The CFT is complemented with appropriate boundary conditions on $P$. A dual gravity  theory will reside in space $N$ with dimension $d + 1$, whose total boundary $\partial N = Q \cup M$ includes a hypersurface $Q$ in the bulk of gravity, such that $\partial Q = \partial M = P$. Such constructions can be realized in full (top-down) string theory, where the role of hypersurface $Q$ of boundary conditions is played by  appropriate $D$-branes~(see \cite{Karch:2000gx,DeWolfe:2001pq,Bak:2003jk,DHoker:2007zhm,DHoker:2007hhe,Aharony:2011yc} for some examples), so in full string theory the hypersurface $Q$ is dynamically fixed.

In the proposal of~\cite{Takayanagi:2011zk} the dynamics of branes is replaced by Neumann boundary condition, which is expected to correctly account for the backreaction of the part of the gravity theory beyond the end-of-the-world $Q$:
\begin{equation}
K_{ab}-Kh_{ab} \ = \ \kappa T_{ab}-\Sigma h_{ab} \,.
\label{Junc-Cond}
\end{equation}
This is in general a second-order equation for the induced metric $h_{ab}$ on $Q$, which defines the embedding of $Q$ in the bulk space $N$ (solving it for $h_{ab}$ is equivalent to solving it for $z(x)$ which determines the embedding of $Q$ in terms of figure~\ref{AdSBCFT}). $K_{ab}$ is the pullback of the extrinsic curvature on $Q$ ($K$ being its scalar). The right hand side of the equation is the stress-energy tensor of the matter placed on $Q$ with units set by $\kappa=8\pi G$, where $G$ being the $d+1$-dimensional  Newton's constant. Part of the stress-energy tensor corresponding to a constant energy density $\Sigma$ is separated. $\Sigma$ may also be referred as to the surface tension, or equivalently, cosmological constant on $Q$.

In the next section we will describe some solutions to equation~(\ref{Junc-Cond}) with $T_{ab}=0$. Since the Neumann boundary condition should reflect the choice of the dual boundary condition on $P$, $\Sigma$ should have a meaning in terms of the CFT data. The exact meaning has not been established so far, but some insight can be obtained by studying the entropy of the defect created by $P$, as in~\cite{Takayanagi:2011zk,Magan:2014dwa,Cavalcanti:2018pta}.

We will restrict our interest to 1+1 CFT examples and to asimptotically $AdS_3$ bulk geometries in the Poincaré patch. We will use $x^\mu=(t,x)$ as the CFT coordinates and $z$ as the coordinate in the gravity bulk.
 
The main player in our game will be the asimptotically anti de Sitter geometry given by the metric
\begin{equation}
ds^{2}=\frac{L^{2}}{z^{2}}\left(
-f(z)dt^{2} + dx^{2} + \frac{dz^{2}}{f(z)}\right)\ .  \label{AdSMetric}
\end{equation}
Pure AdS space is represented by $f(z) = 1$, while the BTZ black hole geometry is obtained when  $f(z)=1-z^{2}/z_{h}^{2}$. The latter geometry is dual to a finite temperature CFT state, with temperature given by $T=1/(2\pi z_{h})$, where $z_h$ is the coordinate of the horizon of the black hole. The two solutions can be related to each other by a (large) diffeomorphism. We will make use of this fact in the next section.

The simplest static solution of boundary condition~(\ref{Junc-Cond}) is obtained for the half-plane configuration. We choose boundary $P$ to be parameterized by equation $x=0$ (as in figure~\ref{AdSBCFT}). The embedding of $Q$ can then be parameterized by $x=x(z)$. For $T_{ab}=0$ and empty $AdS_3$ equation~(\ref{Junc-Cond}) is solved by a straight line embedding~\cite{Takayanagi:2011zk},
\begin{equation}
x\left(z\right) \ = \ z\cot \theta \,, \qquad \text{where} \qquad \cos \theta =L \Sigma .  
\label{7}
\end{equation}
Tension $\Sigma$ defines the angle, at which plane $Q$ intersects the asymptotic AdS boundary. We define $\theta$ as the angle external to region $N$ encoding physics in $M$. It can be seen that the case $0\leq \theta <\pi /2$ corresponds to $\Sigma>0$. The tension is negative for $\pi /2<\theta \leq \pi $. In both cases the tension is bounded: $\left\vert \Sigma \right\vert \leq 1/L$.

In the finite temperature geometry the solution is slightly more involved,
\begin{equation}
x\left( z\right) =z_{h}\text{arsinh}\left( \frac{z}{z_{h}}\cot \theta
\right) \text{ }.  \label{8}
\end{equation}
Angle $\theta$, again, is the angle at which $Q$ crosses the boundary at $z = 0$, external to subspace $N$. In the $z\rightarrow 0$ ($f(z)\to 1$) limit one recovers the pure AdS result (\ref{7}).

Some other solutions to boundary conditions~(\ref{Junc-Cond}) were considered in~\cite{Nozaki_2012,Magan:2014dwa,Erdmenger_2015,Seminara:2017hhh,Seminara:2018pmr,Cavalcanti:2018pta,Shashi:2020mkd,Sato:2020upl}. In the next section we will generate time dependent finite temperature solutions applying a conformal transformation.

\section{Time-dependent AdS/BCFT solutions}
\label{sec:solutions}

Other solutions of equation~(\ref{Junc-Cond}) can be generated by applying isometries to the basic solution (\ref{7}). The AdS$_{d+1}$ metric is invariant under the $d$-dimensional conformal group. For example, boosts create lines $x=z\cot\theta(\eta)$ moving with constant velocity $\eta$ in the $AdS_3$ bulk, intercepting the plane $z=0$ at an $\eta$-dependent angle~\cite{Cavalcanti:2018pta}.

More interesting configurations can be obtained from the special conformal transformations~\cite{Takayanagi:2011zk}. In the Euclidean space ($t \rightarrow it_E$) the half-plane $x>0$ on the boundary can be mapped to the interior of a disc by a global transformation~\cite{Berenstein:1998ij}
\begin{equation}
x_{\mu }^{\prime }=\frac{x_{\mu }+c_{\mu }x^{2}}{1+2\left( c\cdot x\right)
+c^{2}x^{2}} \ , \label{SCTx}
\end{equation}
where $c_{\mu}$ is a constant vector and $x^\mu = (x,t_E)$. The map of the half-plane $x=0$ to the disc of radius $R$ corresponds to the choice $c_{\mu}=(1/2R,0)$. The AdS metric is invariant under this transformation provided the coordinate $z$ is transformed as
\begin{equation}
z^{\prime }=\frac{z}{1+2\left( c\cdot x\right) +c^{2}x^{2}} \ . \label{SCTz}
\end{equation}
In the bulk, the transformation maps the two-dimensional Euclidean AdS$_2$ slices, including $Q$~(\ref{7}) into spherical domes sitting on $M$. The new $Q$ is defined by equation
\begin{equation}
t_{E}^{2}+\left( x-R\right) ^{2}+\left( z-R\cot \theta \right)^2 =R^{2}\csc
^{2}\theta \ . \label{esferaAdS}
\end{equation}
As before, $\theta$ is the external intersection angle of the spherical surface with the $z=0$ boundary. When tension $\Sigma = 0$, or $\theta = \pi /2$, $Q$ is exactly a hemisphere. One can also consider the analytic continuation of the spherical solution to the Minkowski space.
\begin{equation}
-t^{2}+\left( x-R\right) ^{2}+\left( z-R\cot \theta \right)^2 =R^{2}\csc
^{2}\theta \ . \label{hiperboloideAdS}
\end{equation}
The real-time solution describes a compact space with expanding walls. 

\begin{figure}[tbh]
\centering
\includegraphics[height=6cm]{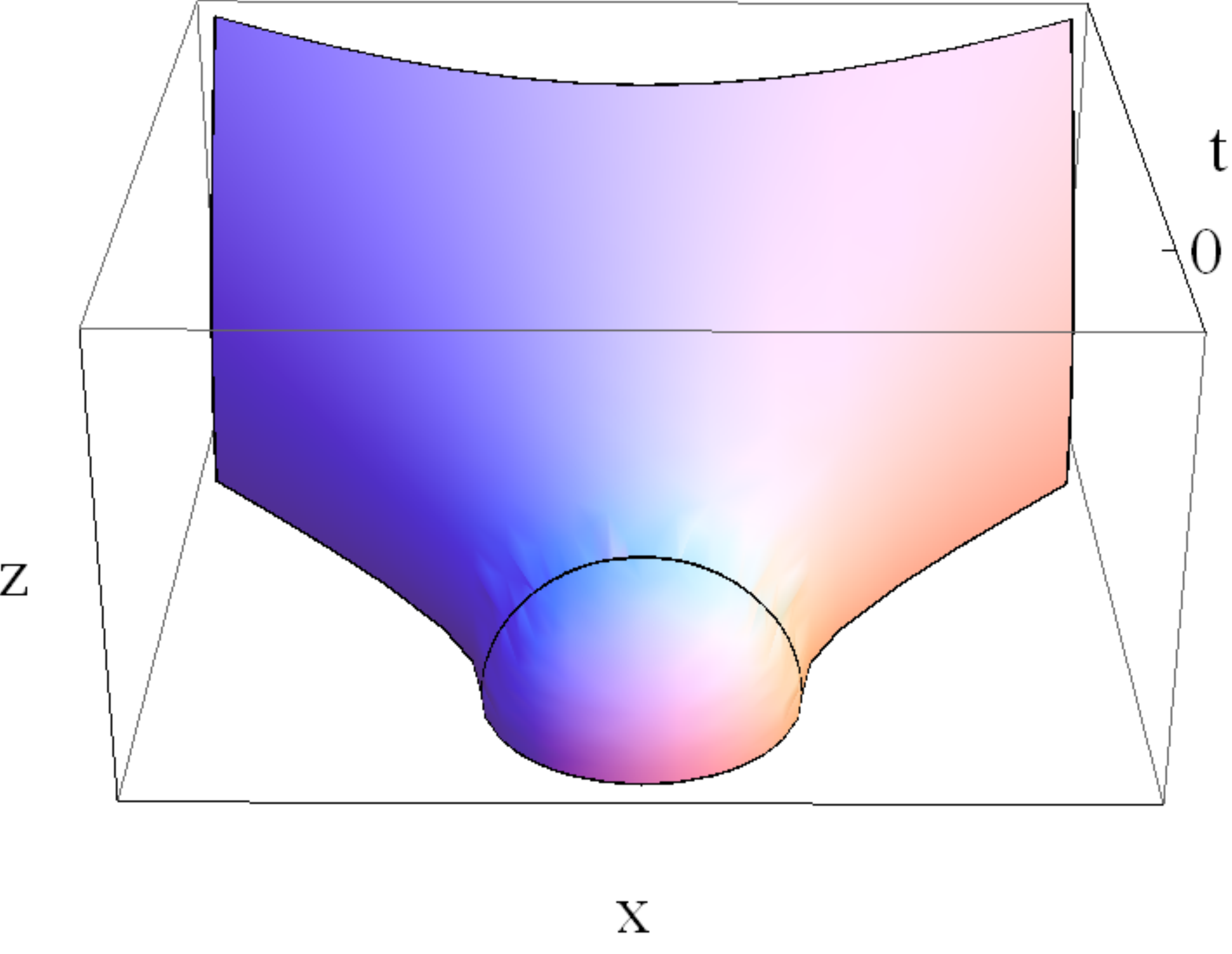}\quad \quad \quad \quad
\caption{Nucleation of a Euclidean bubble of anti-de Sitter space $t<0$, creation ($t=0$) and evolution of the real bubble ($t>0$).} \label{figure1} 
\end{figure}

One possible application of such solutions is in the context of the dynamics of phase transitions, or the problem of the decay of a false vacuum. Euclidean solution~(\ref{esferaAdS}) describes an imaginary time nucleation of a bubble of a new phase, while Minkowskian solution~(\ref{hiperboloideAdS}) -- the expansion of the bubble after the nucleation (figure~\ref{figure1}). 
In this setup, the anti-de Sitter space represents the true vacuum, while the effect of the false vacuum is effectively described through the non-zero surface tension. As we shall see, in this model, temperature effects accelerate the expansion, and there is no finite temperature phase transition.

To generalize the above solutions to the case of finite temperature we are going to apply a bulk diffeomorphism, which relates the empty $AdS_3$ geometry with that of the BTZ black hole. Since equation~(\ref{Junc-Cond}) is a tensor equation, we expect it to transform covariantly under the diffeomorphism. That is, it maps solutions of the equation to other solutions of the equation.

Let us consider a general transformation of the metric
\begin{equation}
g_{\mu \nu }^{\prime }=\frac{\partial x^{\alpha }}{\partial x^{\prime \mu }}
\frac{\partial x^{\beta }}{\partial x^{\prime \nu }}g_{\alpha \beta } \ . 
  \label{MetricTrans}
\end{equation}
To distinguish between AdS and BTZ coordinates in the above equation, we reserve the original coordinates $\left( t,x,z\right)$ for the empty AdS, and use primed coordinates, $\left( t^{\prime },x^{\prime },z^{\prime }\right)$ for the BTZ. With these definitions, we obtain the following coordinate transformation
\begin{eqnarray}
t &=&-z_{h}+\frac{z_{h}^{2}\cosh \left( x^{\prime }/z_{h}\right) \text{e}%
^{t^{\prime }/z_{h}}}{\sqrt{z_{h}^{2}-z^{\prime 2}}} \ ,  \nonumber \\
x &=&\frac{z_{h}^{2}\sinh \left( x^{\prime }/z_{h}\right) \text{e}%
^{t^{\prime }/z_{h}}}{\sqrt{z_{h}^{2}-z^{\prime 2}}} \ , \label{AdS-BTZ-trans} \\
z &=&\frac{z_{h}z^{\prime }\text{e}^{t^{\prime }/z_{h}}}{\sqrt{%
z_{h}^{2}-z^{\prime 2}}} \ . \nonumber
\end{eqnarray}%
In deriving these formulae, the integration constants were fixed by imposing the condition that in the limit  $z_{h}\rightarrow \infty$ we must recover $t=t^{\prime },$ $x=x^{\prime }$ e $z=z^{\prime }$.

Transformations~(\ref{AdS-BTZ-trans}) can be readily used on equation~(\ref{esferaAdS}) to obtain a new, temperature and time dependent solution of the AdS/BCFT boundary condition~(\ref{Junc-Cond}):
\begin{eqnarray}
t_{E}^{\prime }=z_{h}\text{arccos}\left\{ \frac{1}{z_{h}\sqrt{f\left(
z^{\prime }\right) }}\left[ z_{h}\cosh \left( x^{\prime }/z_{h}\right)-R\frac{z^{\prime }}{z_{h}}\cot \theta -R\sinh \left( x^{\prime }/z_{h}\right) 
\right] \right\} .  \label{Esfera-BTZ}
\end{eqnarray}
The analytical continuation to the Minkowski space is performed by doing a Wick rotation $t\rightarrow it_{E}$ and $t'\rightarrow it'_{E}$, so we also find a hyperboloid-like solution
\begin{eqnarray}
t^{\prime }=z_{h}\text{arccosh}\left\{ \frac{1}{z_{h}\sqrt{f\left(
z^{\prime }\right) }}\left[ z_{h}\cosh \left( x^{\prime }/z_{h}\right)-R\frac{z^{\prime }}{z_{h}}\cot \theta -R\sinh \left( x^{\prime }/z_{h}\right) 
\right] \right\} .  \label{Hiperb-BTZ}
\end{eqnarray}
The parameter $R$ in the transformed solutions becomes the height of the profile in the $z$ direction, $0\leq R\leq z_h$. If one, however, analytically continues to $R>z_h$ one would get a class single-boundary solutions, whose  $R\to\infty$ limit is static solution~(\ref{8}). We will not consider this branch here.

\begin{figure}[tbh]
\centering
\includegraphics[height=7cm]{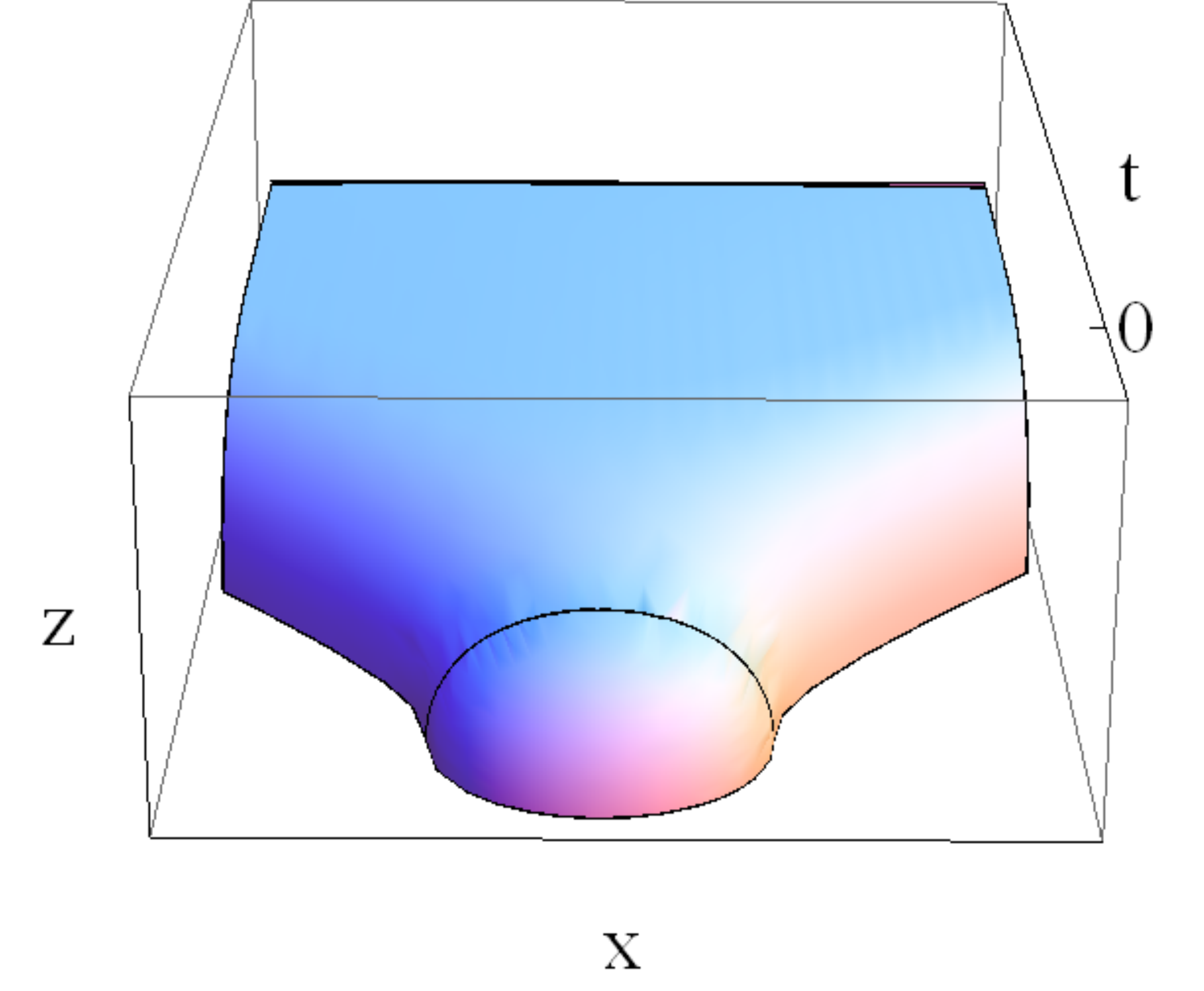}
\caption{Nucleation of a Euclidean bubble of the BTZ spacetime $t<0$. Creation ($t=0$) and evolution of the real bubble ($t>0$), for $\theta = \pi /2$.} \label{figure2} 
\end{figure}

The characteristic shape of the hypersurface $Q$ is demonstrated by figure~\ref{figure2}, where again, the Euclidean solution is glued with the Minkowskian one at $t=0$. In comparison with configuration shown on figure~\ref{figure1}, the new $Q$ is bounded in $z$ dimension by the horizon of the black hole. One can also check that (\ref{Esfera-BTZ}) and (\ref{Hiperb-BTZ}) recover shapes~(\ref{esferaAdS}) and (\ref{hiperboloideAdS}) respectively, in the $z_h\rightarrow\infty$ limit.

\begin{figure}[tbh]
\centering
\includegraphics[height=6cm]{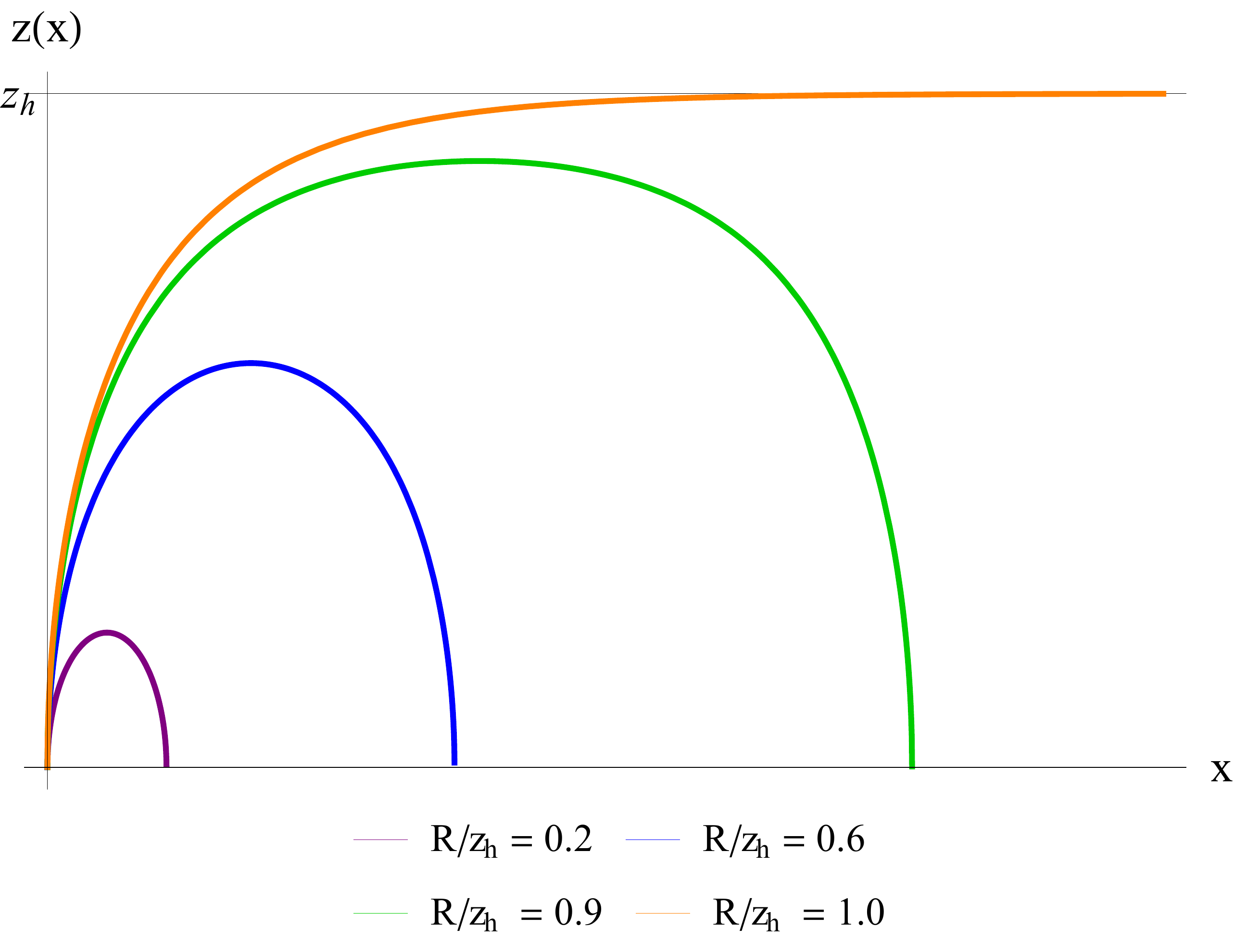}
\caption{Profiles of $Q$ in the BTZ geometry for different values of $R$ (height of the bubble), at $t = 0$ and $\theta =\pi/2$. As $R$ approaches the horizon, the width of the bubble becomes infinite, $l=2\rho\to\infty$.} \label{figure3} 
\end{figure}

In the finite temperature case the parameter $R$ is related to the spatial radius $\rho$ at time $t=0$ through
\begin{eqnarray}
\rho 
\ = \ z_h{\rm arctanh}\frac{R}{z_h} \,,
\label{bubblesize}
\end{eqnarray}
which makes sense only when $R\leq z_h$, and for $z_h\to \infty$ reduces to $\rho=R$. We also see that the profile intercepts the horizon at infinite radius $\rho\rightarrow\infty$, or $R \rightarrow z_{h}$. The profiles of $t=0$ configurations is shown on figure~\ref{figure3}.

\begin{figure}[tbh]
\includegraphics[height=0.3\linewidth]{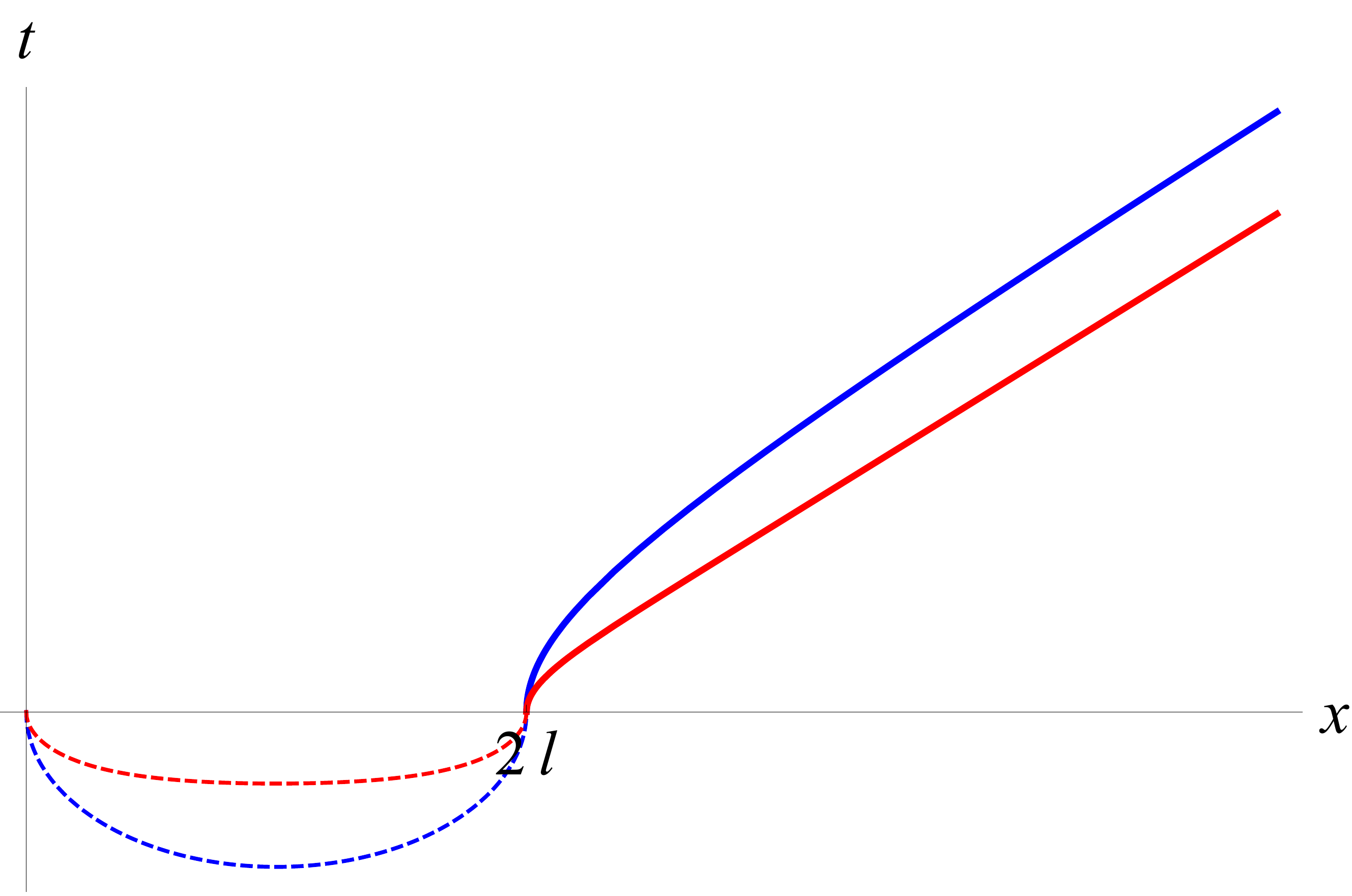} 
\hfill{
\includegraphics[height=0.3\linewidth]{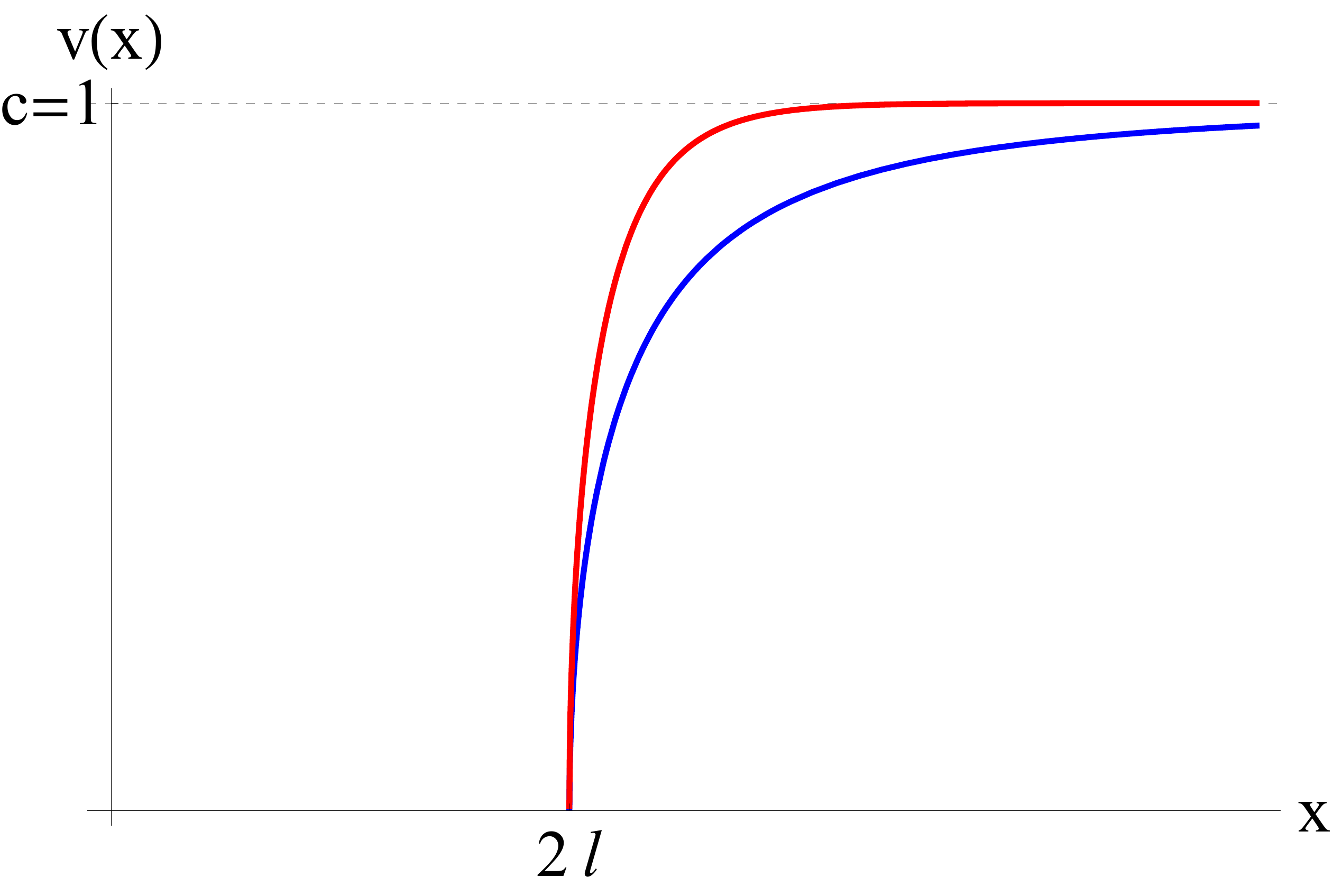}
}
\caption{(Left) trajectories of the $T=0$ (\ref{hiperboloideAdS}) (blue curve) and $T\neq 0$ (\ref{Hiperb-BTZ}) (red curve) bubble walls. (Right) velocities of the two types of walls as a function of $x$.} 
\label{figura5}
\end{figure}

It is also interesting to compare the rate of expansion of the bubbles at zero and finite temperature. On figure~\ref{figura5}, we show the trajectories of walls of two bubbles in zero (blue) and non-zero (red) temperature geometries, as well as the velocities of the walls. Both bubbles have the same size at the moment of creation and their walls are at rest. At late times the walls of either bubble approach the speed of light. However, at finite temperature the walls have larger acceleration. Although the plots are shown for $\Sigma=0$, this effect does not depend on the tension. Consequently, in this model one does not observe a critical (stationary) bubble at any temperature.

\section{Local quantum quench}
\label{sec:examples}

The time-dependent solutions of the AdS/BCFT problem can also be discussed in a more general context of the dynamics of quantum systems out of equilibrium. One can apply an abrupt local, or global perturbation (quench) to the system and study the time behavior of the correlators. Entanglement entropy is a measure of quantum correlations. For this reason it is an interesting object to study in the non-equilibrium dynamics.

One distinguishes global and local quenches. In global quenches one perturbs the system as a whole, for example, by tuning its Hamiltonian. In local quenches, one only perturbs a part of the system. A well-established result in the case of global quenches is the linear growth of the entanglement entropy~\cite{Calabrese:2005in}. The linear growth saturates for a finite system and for late times, the entropy is constant. These observations are true also beyond the conformal case~\cite{Calabrese:2016xau}.

We would like to compare the AdS/BCFT bubbles constructed in the previous section with local quenches. In \emph{cut and glue} quench protocol two complimentary subsystems are prepared unentangled in their respective ground states. Then the two subsystems are brought together and their joint evolution is investigated. Alternatively, in such a quench, the global system is disconnected at certain moment of time and the dynamics of the isolated subsystems is watched. In $1+1$ CFT this quench protocol can be treated by appropriate conformal transformations, mapping to a simple BCFT configuration.

From the holographic point of view, our prescription is close to the double 
quenches considered in~\cite{Caputa:2019avh}, where two slits in the initial 
state correspond to the walls of our bubble. Here the walls of the bubble are 
entangled through the Euclidean nucleation protocol (see a more recent 
model~\cite{Akal:2020wfl}, where the bubble does not reconnect with the AdS 
boundary). Also, in our analysis, the 
exterior of the bubble will always remain unentangled and disconnected. In such 
a case the problem is solved without invoking a conformal map.

In terms of the holographic Ryu-Takayanagi prescription~\cite{Ryu:2006bv}, which in our case coincides with the more general covariant one of Hubeny, Rangamani and Takayanagi~\cite{Hubeny:2007xt}, either of the two bubbles~(\ref{esferaAdS}), or~(\ref{Esfera-BTZ}) at $t=0$, have zero entanglement with the exterior, as in the cut and glue protocol. This is because by the prescription, the entanglement entropy is proportional to the area (length in 3D) of the minimal area spacelike hypersurface (geodesic line in 3D) $\gamma_{\min}$ that connects the endpoints (walls) of the bubble in the gravity bulk, 
\be
\label{RTeq}
S_{\rm E }\ = \ \frac{{\rm Area}[\gamma_{\min}]}{4G}\,.
\ee
However, in the presence of the end-of-the-world brane $Q$ the geodesic is allowed to end on $Q$. Since $Q$ also ends on the walls of the bubble, the geodesic has zero length and the entanglement entropy is zero. We will illustrate this argument in an example that follows. See also~\cite{Ugajin:2013xxa,Erdmenger_2015,Erdmenger_2016entang,Erdmenger_2016holog,Miao_2017,Seminara:2017hhh,Cavalcanti:2018pta} for a similar discussion.

During the expansion, for $t>0$, the bubble will continue having zero entanglement entropy. By analogy, with quantum quenches, the walls of the bubble follow propagation of the front of the quasiparticles, and there is no entanglement with anything outside of the bubble. (The walls will eventually form a light cone, analogous to the lightcone of the quenched systems~\cite{calabrese2006time,Calabrese:2007rg}.) Below we will focus on the entanglement of other subsystems and compute their entropy.

\begin{figure}[tbh]
\includegraphics[height=0.23\linewidth]{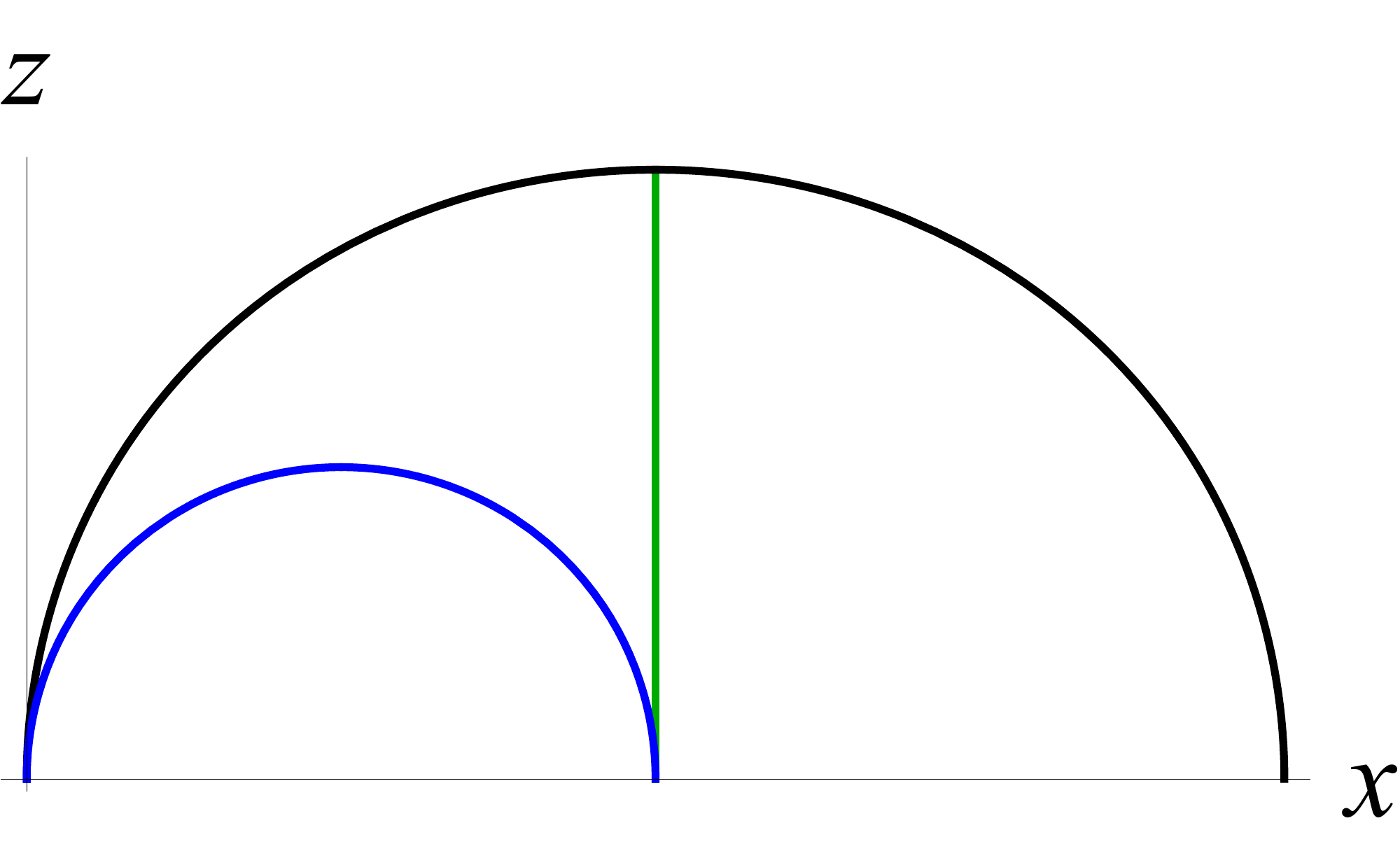} 
\hfill{
\includegraphics[height=0.23\linewidth]{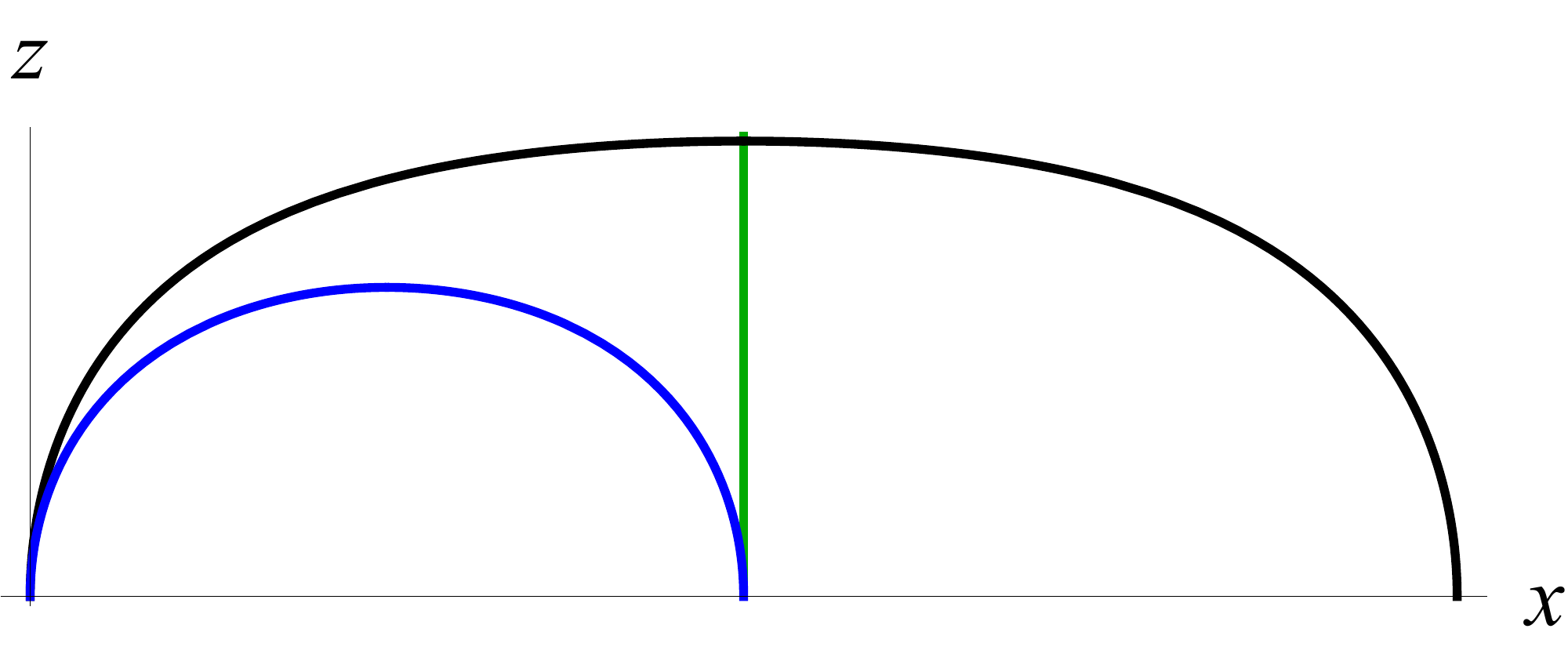}
}
\caption{(Left) geodesics (blue and green lines) and end-of-the-world surface $Q$ (black) in the empty AdS geometry, $\theta = \pi /2$. (Right) same in the BTZ geometry.}
\label{setup}
\end{figure}

The first configuration we are going to analyze is shown on figure~\ref{setup}: we would like to study the entanglement of the two halves of the bubble.  In the figure, the black curves correspond to boundary $Q$ in AdS (left) and BTZ (right) spaces, given by equations (\ref{hiperboloideAdS}) and (\ref{Hiperb-BTZ}), respectively. The profiles with $\theta = \pi /2$ at $t = 0$ are shown. 

Using equation~(\ref{RTeq}) we compute the entanglement entropy of any of the halves of the bubble as the length of a geodesic line connecting the walls of the bubble in the gravity bulk. There are two options in this case (shown as blue and green lines on figure~\ref{setup}): one line connects the endpoints of the interval representing a half of the bubble, while the other option connects the center of the bubble with the curve $Q$. It turns out that the second geodesic (green vertical line on figure~\ref{setup}) is always shorter in either geometry.

In the case of empty AdS (left panel on figure~\ref{setup}) the boundary $Q$ is a circular arc of radius $R\csc\theta$ centered at $z=R\cot\theta$, equation~(\ref{hiperboloideAdS}). For the bubble of radius $r(t)$ at time $t$, $r(t)^{2}=R^2 + t^2$. Calculation then yields the length of the geodesic in geometry~(\ref{AdSMetric}) connecting the middle of the bubble with $Q$ as 
\begin{equation}
    \ell_0 \ = \ L\int _{\epsilon}^{z_\ast} \frac{dz}{z} \ = \ L\log \left(\frac{z_\ast}{\epsilon}\right) \ = \ L\log \left(\frac{\sqrt{R^2\csc ^2\theta + t^2}+R\cot\theta}{\epsilon}\right)\,,
    \label{ComVertADS}
\end{equation}
where $z_\ast$ is the height of the arc $Q$. As usual, $\epsilon$ is a UV cut-off introduced to make the length finite. 

\begin{figure}[tbh]
\includegraphics[height=0.3\linewidth]{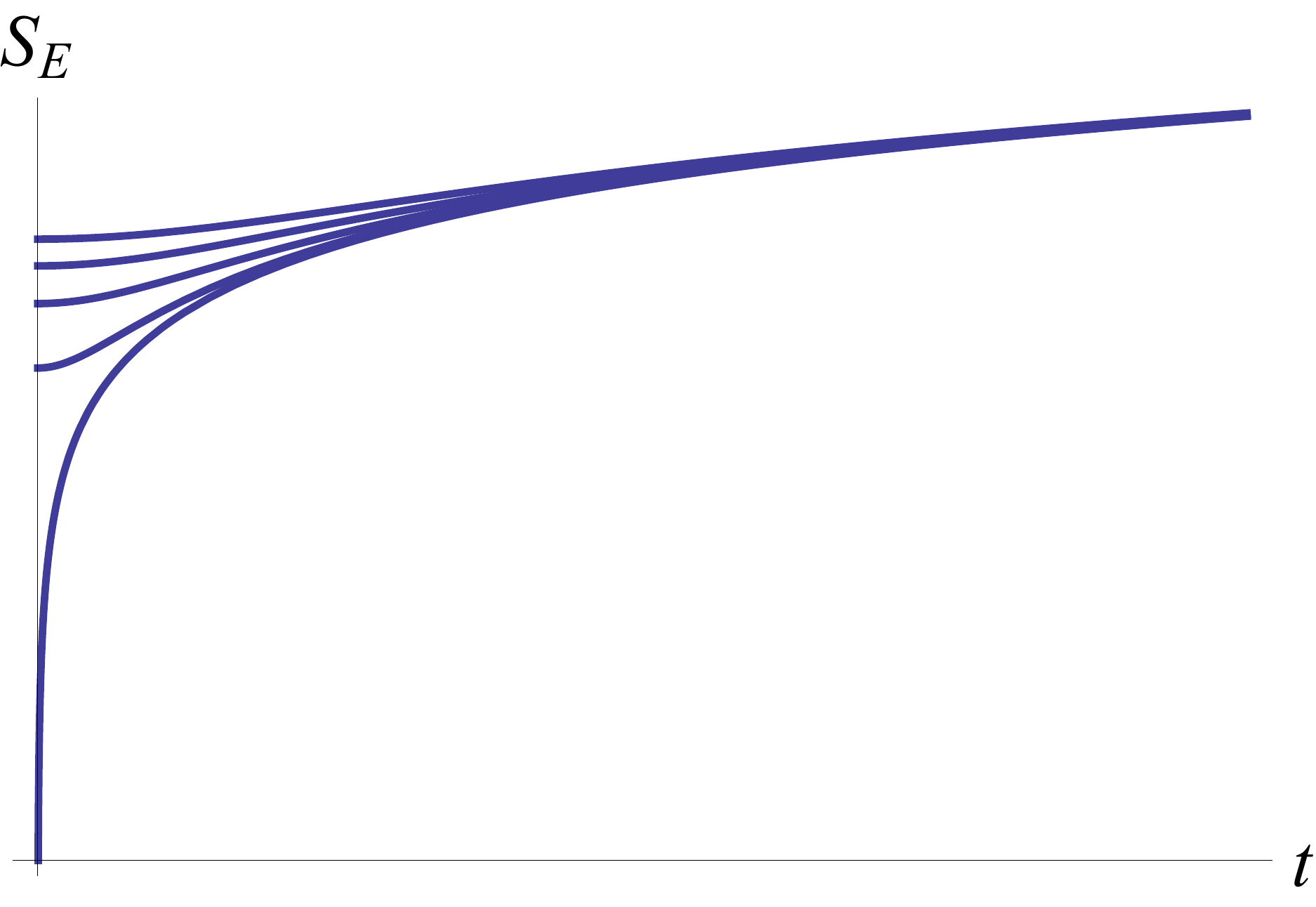}
\hfill{
\includegraphics[height=0.3\linewidth]{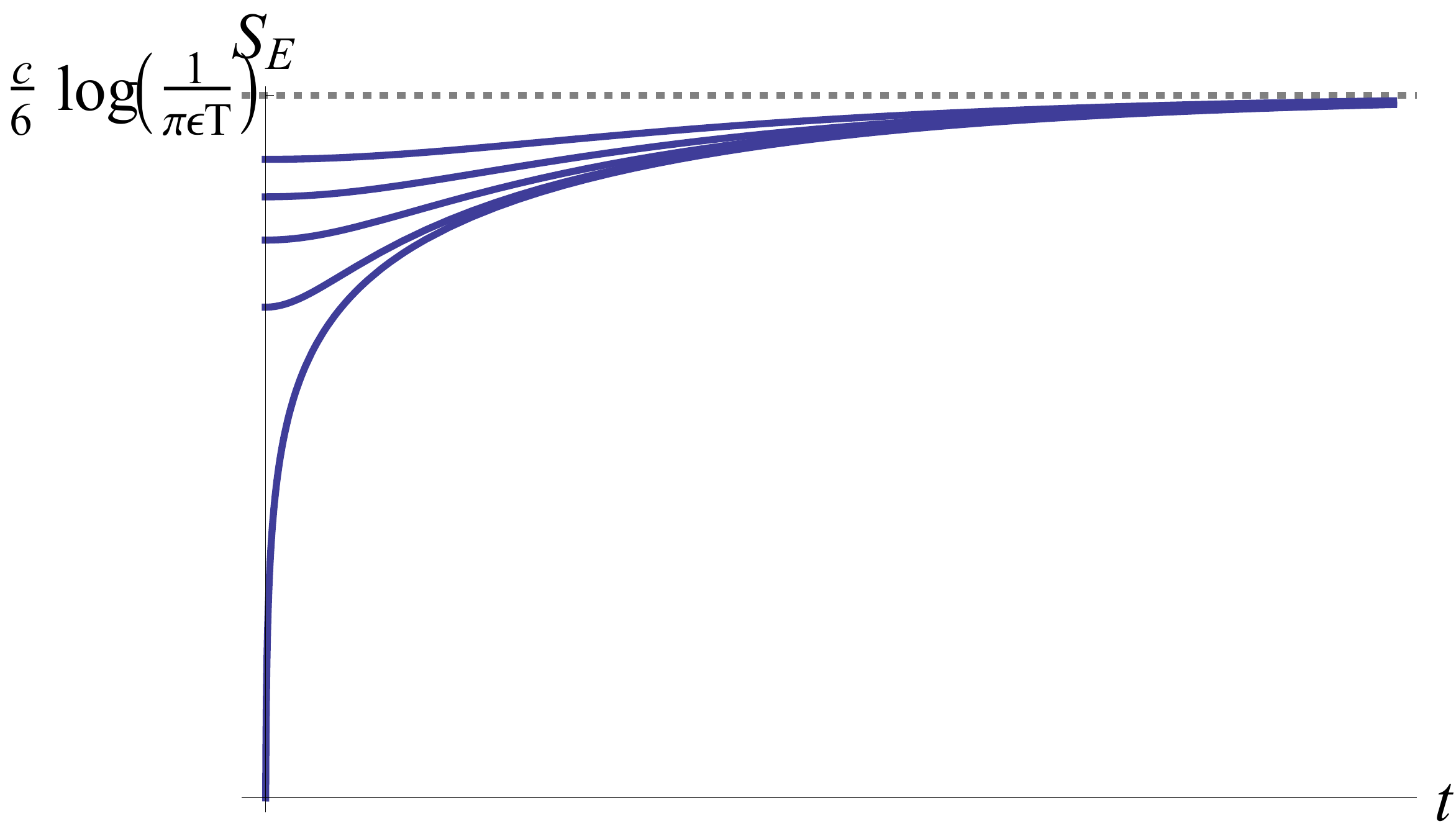}
}
\caption{Plot of the time dependence of the entanglement entropy of the two halves of the expanding bubble for different initial sizes of the bubble. The left plot corresponds to empty AdS geometry shows asymptotic logarithmic growth. The right plot, for the finite temperature geometry, shows saturation of the entropy at a value independent from the initial parameters of the bubble.}
\label{ComprimentoDaLinhaReta}
\end{figure}

Meanwhile the length of the geodesic connecting the two endpoints of the interval of length $r(t)$ (blue circle on figure~\ref{setup}) is given by
\begin{equation}
\ell \ = \ 2\times L\int _{\epsilon}^{r/2} \frac{rdz}{z\sqrt{r^2-4z^2}} \ = \ 2L \log \left(\frac{\sqrt{R^2 + t^2}}{\epsilon}\right)+O(\epsilon^2).
\end{equation}
By choosing the cutoff $\epsilon$ sufficiently small one can always make $\ell>\ell_0$. The reason for that is that the geodesic $\ell_0$ has only one endpoint on the boundary $z=0$, where the distance must be regulated. Therefore we use $\ell_0$ in equation~(\ref{RTeq}). The standard conversion between the 3D gravity and CFT parameters is~\cite{Brown:1986nw}
\be
c \ = \ \frac{3L}{2G}\,.
\ee
At late times the entanglement entropy of a half of the bubble behaves as
\be
S_{\rm E} \ \sim \ \frac{c}{6}\log\frac{t}{\epsilon}\,.
\ee
On figure~\ref{ComprimentoDaLinhaReta} (left) we plot the behavior of the entropy for different values of the initial bubble size. The entropy interpolates between the initial value, cf.~\cite{Cavalcanti:2018pta}
\be
\frac{c}{6}\log \left(\frac{2R}{\epsilon}\cot\frac{\theta}{2}\right)
\ee
controlled by $R$ and $\theta$ and the late time logarithmic growth, independent from those parameters.

A similar calculation can be done in the finite temperature case, as on the right of figure~\ref{setup}. For the qualitative understanding it is sufficient to consider the case $\theta = \pi /2$. As for zero temperature, the geodesic that connects the center of the bubble with $Q$ (green line on figure~\ref{setup}) can always be made shorter if $\epsilon$ is sufficiently small. The turning-point $z_{*}$ (the height) of the $Q$ profile is given by the expression, cf.~(\ref{Hiperb-BTZ})
\begin{equation}
   z_{*} \ = \ \text{sech}(t/z_h) \ \sqrt{z_h^2\text{sech}^2 (t/z_h) +R^2} ~.
\end{equation}
Therefore, the temporal behavior of geodetic $\ell_0$ is
\begin{equation}
    \ell_{0}\ =\ L\int_{\epsilon }^{z_{\ast }}\frac{dz}{z\sqrt{1-\frac{%
z^{2}}{z_{h}^{2}}}}\ =\ L\log \left[ \frac{2z_{h}}{\epsilon }\frac{\text{sech}%
\left( t/z_{h}\right) \sqrt{R^{2}+z_{h}^{2}\sinh ^{2}\left( t/z_{h}\right) }%
}{z_{h}+\sqrt{z_{h}^{2}-\text{sech}^{2}\left( t/z_{h}\right) \left[
R^{2}+z_{h}^{2}\sinh ^{2}\left( t/z_{h}\right) \right] }}\right] ~.
    \label{ComVertBTZ}
\end{equation}
For $t\rightarrow\infty$, the length $\ell_0$ tends to an asymptotic value equal to $L\log(2z_h/\epsilon)$, as can be also seen on figure~\ref{ComprimentoDaLinhaReta} (right). The asymptotic value of the entropy is
\be
\label{finiteTentropy}
S_{\rm E} \to \frac{c}{6}\log\left(\frac{1}{\pi\epsilon T}\right)\,.
\ee
As in the $T=0$ case, the values of $R$ and $\theta$ only affect the early time behavior of the entropy, for example, the $R$-dependence at $t=0$ is given by
\be
S_{\rm E} \ \sim \  \frac{c}{6}\log \left[\frac{1}{\pi\epsilon T}\frac{\sinh2\pi Tl}{1+\cosh2\pi Tl}\right]\,,
\ee
when $t\ll 1/T$. Here we used relation~(\ref{bubblesize}) between parameter $R$ and the initial radius $\rho\equiv l$ of the bubble.

A more common configuration to study in the context of local quenches is the evolution of the entanglement of the interval, corresponding to the $t=0$ bubble with the remainder of the system. Recall that at time $t=0$, the bubble is unentangled with the exterior. We will consider $\theta=\pi/2$ and first treat the case of zero temperature.

\begin{figure}[t]
\centering
\includegraphics[height=5.5cm]{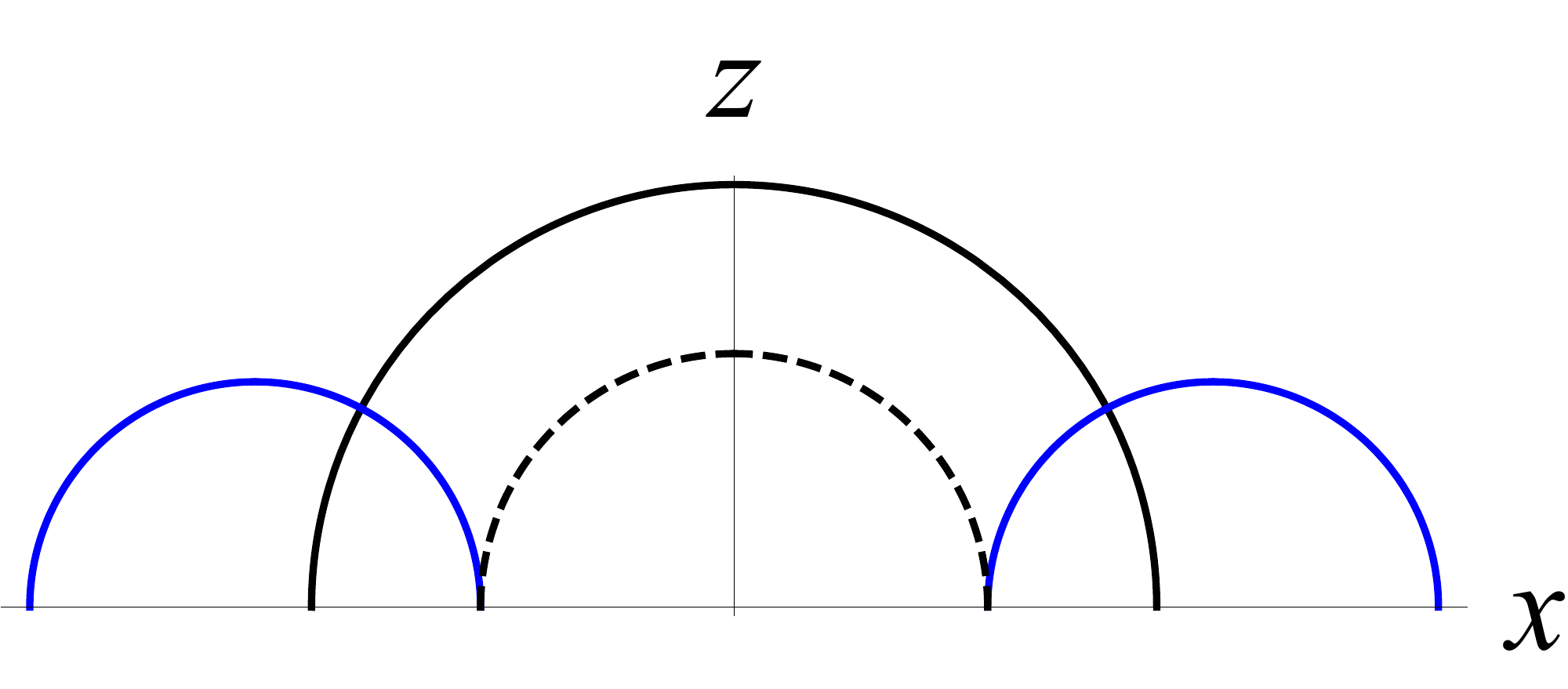} 
\caption{Calculation of the entanglement entropy of the finite interval bounded by the dashed arc with the rest of the system after a cut and glue quench.}
\label{SetupDiferente}
\end{figure}

The setup we are going to study is shown on figure~\ref{SetupDiferente}. The dashed black arc of radius $R$ is the $Q$-profile of the initial bubble created at $t = 0$. The bubble begins to expand, and the profile of $Q$ at some later time $t$ is shown as a continuous black arc on figure~\ref{SetupDiferente}. At this time the radius of $Q$ (and the radius of the bubble) is given by $l=\sqrt{t^2+R^2}$. 

We would like to know the entanglement of the initial interval of size $l=2R$ with the rest of the system. For this we are going to use equation~(\ref{RTeq}) with an appropriate minimal geodesic line. The blue curves of radius $r$ on figure~\ref{SetupDiferente} correspond to the natural choice of the minimal surface at early times, after the beginning of the expansion. An alternative choice would be a geodesic connecting the endpoints of the interval, which in this case coincides with the dashed line. Clearly the second geodesic is longer. The two blue pieces have zero length at $t=0$ and the entanglement is zero as claimed in the beginning of this section.  

In order to apply equation~(\ref{RTeq}) we have to calculate the length of the blue curves, which is inside boundary $Q$, from the endpoints of the interval until the intersection point $z_{0}$ with the black curve. The geodesics must satisfy Dirichlet  boundary condition at the endpoints and they must be perpendicular to the black curve at the intersection point (Neumann boundary conditions). The blue geodesics are also circular arcs with 
\begin{equation}
    r=\frac{l^2-R^2}{2R} \qquad \text{and} \qquad z_0 = \frac{r\sqrt{R(R+2r)}}{R+r}~.
\end{equation}
Therefore, the length of the relevant pieces of the geodesics is
\begin{equation}
    \ell \ = \ 2\times L\int _{\epsilon}^{z_0}\frac{dz}{z}\frac{r}{\sqrt{r^2-z^2}} 
    \ = \ 2L\log \left(\frac{t^2}{\epsilon\sqrt{t^2+R^2}}\right) + O(\epsilon^2)\,. 
    \label{9}
\end{equation}

However, as the continuous black arc expands over time, the blue geodesic also expands in size so that there will be a certain instant of time that its length $\ell$ is greater than the length of the dashed black curve $\ell_0$. At this moment we have to switch to $\ell_0$ in equation~(\ref{RTeq}). In this model the change is non-analytic. The length of the dashed black circle is given by
\begin{equation}
    \ell_0 \ = \ 2\times L\int _{\epsilon}^{R}\frac{dz}{z}\frac{R}{\sqrt{R^2-z^2}} \ = \ 2L\log \left(\frac{2R}{\epsilon}\right) + O(\epsilon^2)\,.
    \label{10}
\end{equation}
The phase transition occurs at $t_c = R\sqrt{2(1+\sqrt{2})}$. 

\begin{figure}[thb]
\includegraphics[height=0.36\linewidth]{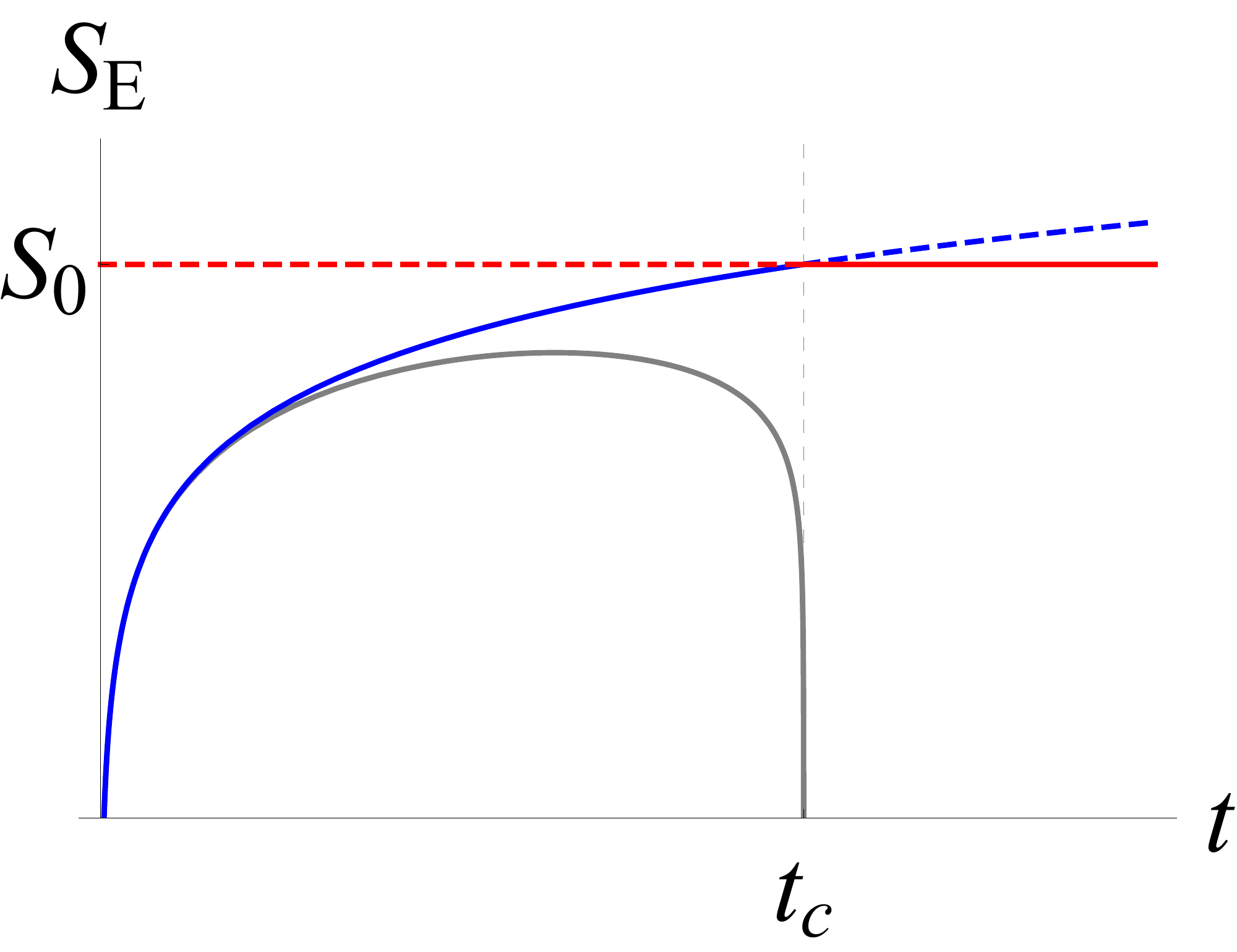}
\hfill{
\includegraphics[height=0.36\linewidth]{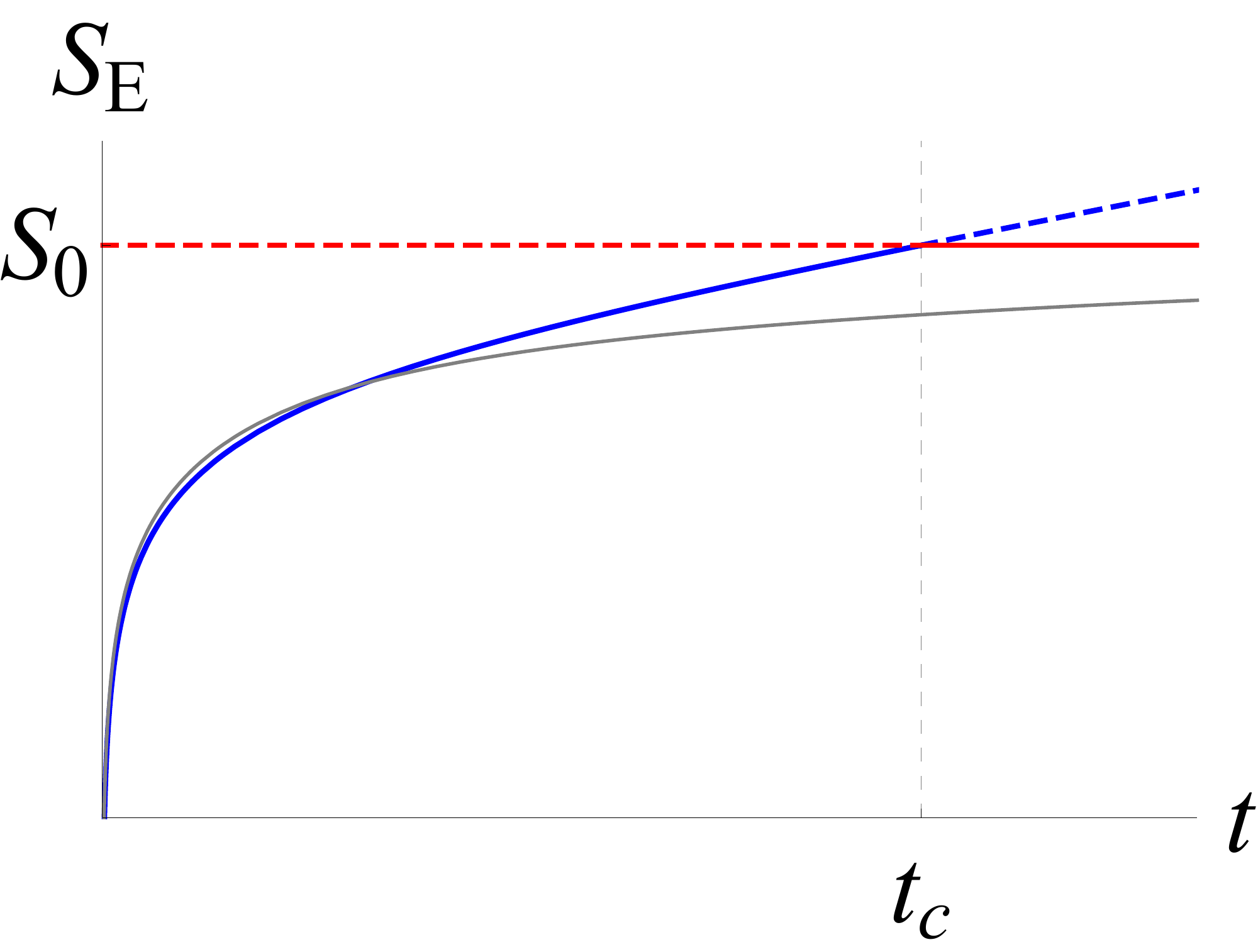}
}
\caption{(Left) Evolution of the entanglement entropy of a finite interval in a bubble quench from the AdS/BCFT calculation (configuration of figure~\ref{SetupDiferente}). The blue curve shows the logarithmic growth from equation (\ref{9}). The growth saturates at $S_0$ (red line), equation (\ref{10}). The gray line is the behavior predicted in a local cut and glue quench, equation~(\ref{logsine}) (Right) Similar plot for $T\neq 0$. The blue line has a linear segment described by equation~(\ref{lineargrowth}). The gray line illustrates the $T=0$ curve.}
\label{PhaseTransition}
\end{figure}

At initial times the entropy of the system grows logarithmically,
\be
\label{CAGentropy}
S_{\rm E} \ \sim \ \frac{c}{3}\log\frac{t^2}{\epsilon R}\,, \qquad t\ll R\,.
\ee
At later times it saturates at the standard universal value of the entropy of a finite interval. The plot of the entropy and of the phase transition is shown on figure~\ref{PhaseTransition}~(left).

{To find the entropy at finite temperature one can apply transformations~(\ref{AdS-BTZ-trans}). Consequently, one derives the following analytical formula}
\begin{equation}
S_E=\frac{c}{3}\log \left[ \frac{2z_{h}\left( \cosh \left( t/z_{h}\right) -1\right) \text{%
e}^{t/z_{h}}}{\epsilon \sqrt{2\left( \cosh \left( t/z_{h}\right) -1\right) 
\text{e}^{t/z_{h}}+\sinh ^{2}\left( l/2z_{h}\right) \text{e}^{2t/z_{h}}}}%
\right],   \label{A}
\end{equation}
where $l$ is the length (rather than radius) of the initial bubble. For short times the generalization of equation~(\ref{CAGentropy}) is
\begin{equation}
S_E\sim \frac{c}{3}\log \left[ \frac{%
2\pi T}{\epsilon }\frac{t^{2}}{\sinh \left( \pi Tl\right)} \right] \,.
\label{tpequenoBTZ}
\end{equation}

It is interesting that at intermediate times, $T^{-1}\ll t <l$, the entropy grows linearly,
\be
\label{lineargrowth}
S_E\sim \frac{c}{3}2\pi T\left(t-\frac{l}{2}\right) + \frac{c}{3}\log\frac{1}{\pi\epsilon T}\,.
\ee
We note that the entropy is twice the value of equation~(\ref{finiteTentropy}) at $t=l/2$. 

For $t>l$ one expects the saturation phase transition to the value of the entropy of an interval at finite temperature:
\be
S_0 \ = \ \frac{c}{3}\log\left[\frac{1}{\pi T\epsilon}\sinh\left(2\pi Tl\right)\right]\,.
\ee
This finite temperature behavior is illustrated by the right plot of figure~\ref{PhaseTransition}.

\section{Conclusions}
\label{sec:conclusions}

In this paper we obtained new time-dependent solutions to the AdS/BCFT problem of~\cite{Takayanagi:2011zk}. These new solutions correspond to expanding walls in anti-de Sitter space and are natural to discuss in the context of the process of nucleation and expansion of new phases.

We demonstrated that the time-dependent solutions can be suitable for the 
discussion of quenched dynamics, in a setup similar to the local cut and glue 
protocol. This observation is supported by the behavior of the entanglement 
entropy, which grows logarithmically at early times. However, our protocol is 
slightly different from the conventional cut and glue quench, and the late 
behavior of the entropy is different.

In the bubble protocol one can observe other characteristic features of quenched dynamics. The solutions are compatible with the light-cone expansion of the correlations. The entanglement of a finite interval saturates at finite time. This effect is typically discussed in the case of the global quench~\cite{Calabrese:2005in}, and since we disregard the exterior of the bubble, it is perhaps not so surprising that the linear behavior also occurs in the bubble quench. 

Some analytical results for local quenches of finite intervals are harder to obtain using the CFT techniques~\cite{Calabrese:2007rg}, because the appropriate conformal maps become non-invertible. Similar problems can occur in the geometric analysis, although we have been able to get some exact analytical results. In particular, at early times we are able to derive equations~(\ref{9}) and~(\ref{CAGentropy}) for the entropy. The early time behavior is compatible with
\be
\label{logsine}
S_{\rm E} \ = \ \frac{c}{3} \log\left(\frac{2t}{\pi\epsilon}l\sin\frac{\pi t}{l}\right)+k'\,.
\ee
which is the cut and glue quench result in 
CFT~\cite{Calabrese:2007mtj,2008JSMTE..01..023E} and in the holographic 
models~\cite{Ugajin:2013xxa,Asplund:2013zba,Asplund:2014coa,Mandal:2016cdw,
Shimaji:2018czt, Caputa:2013eka,Ageev:2019fjf,Kudler-Flam:2020url}. First, the 
match is up to the factor of the length $l$. Naively, the argument of the 
logarithm in equation~(\ref{logsine}) is not dimensionless, which means that 
there is a dimensionful scale hidden in the non-universal part $k'$. The 
holographic derivation automatically gives the correct dimension removing one 
$l$ factor. The saturation point in the bubble quench occurs at $t\sim 1.05 l$, 
which is very close to the point, where the argument of the logarithm 
in~(\ref{logsine}) vanishes, as can also be seen on figure~\ref{PhaseTransition} 
(left).

Finally we derive equations for the evolution of the entropy of a finite interval at finite temperature. For early times we derive equation~(\ref{tpequenoBTZ}), which could be tested by CFT techniques, cf.~\cite{Wen:2018svb}. As already mentioned, at finite $T$, the entropy tends to show linear growth~(\ref{lineargrowth}) for intermediate times $T^{-1}\ll t < l$, before the saturation phase transition. 

We believe that some of the outstanding issues of the present analysis, as well as other interesting questions of the quenched dynamics can be further addressed in the AdS/BCFT formalism. We leave this for a future work. One important question, is whether one can make the AdS/BCFT correspondence more precise by relating $\theta$ (or $\Sigma$) to CFT quantities, or finding appropriate forms of $T_{ab}$ in equation~(\ref{Junc-Cond}).

\paragraph*{Acknowledgements.} We would like to thank Madson R.~O.~Silva for 
collaboration on parts of this project. We also grateful to J\'er\^ome Dubail 
and, especially, to Zixia Wei for useful correspondence regarding the first 
version of the paper. The work of AC was supported by the Brazilian Ministry of 
Education (MEC), the work of DM was supported by the Russian Science Foundation 
grant No.~16-12-10344.

\bibliographystyle{hieeetr}
\bibliography{references}

\begin{thebibliography}{10}

\bibitem{di1996conformal}
P.~Di~Francesco, P.~Mathieu, and D.~S{\'e}n{\'e}chal, {\em Conformal Field
  Theory}.
\newblock Graduate texts in contemporary physics, Island Press, 1996.

\bibitem{polkovnikov2011colloquium}
A.~Polkovnikov, K.~Sengupta, A.~Silva, and M.~Vengalattore, ``Colloquium:
  Nonequilibrium dynamics of closed interacting quantum systems,'' {\em Reviews
  of Modern Physics}, vol.~83, no.~3, p.~863, 2011.

\bibitem{eisert2015quantum}
J.~Eisert, M.~Friesdorf, and C.~Gogolin, ``Quantum many-body systems out of
  equilibrium,'' {\em Nature Physics}, vol.~11, no.~2, pp.~124--130, 2015.

\bibitem{d2016quantum}
L.~D'Alessio, Y.~Kafri, A.~Polkovnikov, and M.~Rigol, ``From quantum chaos and
  eigenstate thermalization to statistical mechanics and thermodynamics,'' {\em
  Advances in Physics}, vol.~65, no.~3, pp.~239--362, 2016.

\bibitem{gogolin2016equilibration}
C.~Gogolin and J.~Eisert, ``Equilibration, thermalisation, and the emergence of
  statistical mechanics in closed quantum systems,'' {\em Reports on Progress
  in Physics}, vol.~79, no.~5, p.~056001, 2016.

\bibitem{altman2015non}
E.~Altman, ``Non equilibrium quantum dynamics in ultra-cold quantum gases,''
  {\em arXiv preprint arXiv:1512.00870}, 2015.

\bibitem{Calabrese:2016xau}
P.~Calabrese and J.~Cardy, ``{Quantum quenches in 1 + 1 dimensional conformal
  field theories},'' {\em J. Stat. Mech.}, vol.~1606, no.~6, p.~064003, 2016,
  1603.02889.

\bibitem{greiner2002collapse}
M.~Greiner, O.~Mandel, T.~W. H{\"a}nsch, and I.~Bloch, ``Collapse and revival
  of the matter wave field of a bose--einstein condensate,'' {\em Nature},
  vol.~419, no.~6902, pp.~51--54, 2002.

\bibitem{trotzky2012probing}
S.~Trotzky, Y.-A. Chen, A.~Flesch, I.~P. McCulloch, U.~Schollw{\"o}ck,
  J.~Eisert, and I.~Bloch, ``Probing the relaxation towards equilibrium in an
  isolated strongly correlated one-dimensional bose gas,'' {\em Nature
  physics}, vol.~8, no.~4, pp.~325--330, 2012.

\bibitem{cheneau2012light}
M.~Cheneau, P.~Barmettler, D.~Poletti, M.~Endres, P.~Schau{\ss}, T.~Fukuhara,
  C.~Gross, I.~Bloch, C.~Kollath, and S.~Kuhr, ``Light-cone-like spreading of
  correlations in a quantum many-body system,'' {\em Nature}, vol.~481,
  no.~7382, pp.~484--487, 2012.

\bibitem{bloch2012quantum}
I.~Bloch, J.~Dalibard, and S.~Nascimbene, ``Quantum simulations with ultracold
  quantum gases,'' {\em Nature Physics}, vol.~8, no.~4, pp.~267--276, 2012.

\bibitem{blatt2012quantum}
R.~Blatt and C.~F. Roos, ``Quantum simulations with trapped ions,'' {\em Nature
  Physics}, vol.~8, no.~4, pp.~277--284, 2012.

\bibitem{meinert2013quantum}
F.~Meinert, M.~J. Mark, E.~Kirilov, K.~Lauber, P.~Weinmann, A.~J. Daley, and
  H.-C. N{\"a}gerl, ``Quantum quench in an atomic one-dimensional ising
  chain,'' {\em Physical review letters}, vol.~111, no.~5, p.~053003, 2013.

\bibitem{langen2013local}
T.~Langen, R.~Geiger, M.~Kuhnert, B.~Rauer, and J.~Schmiedmayer, ``Local
  emergence of thermal correlations in an isolated quantum many-body system,''
  {\em Nature Physics}, vol.~9, no.~10, pp.~640--643, 2013.

\bibitem{islam2015measuring}
R.~Islam, R.~Ma, P.~M. Preiss, M.~E. Tai, A.~Lukin, M.~Rispoli, and M.~Greiner,
  ``Measuring entanglement entropy in a quantum many-body system,'' {\em
  Nature}, vol.~528, no.~7580, pp.~77--83, 2015.

\bibitem{calabrese2006time}
P.~Calabrese and J.~Cardy, ``Time dependence of correlation functions following
  a quantum quench,'' {\em Physical review letters}, vol.~96, no.~13,
  p.~136801, 2006.

\bibitem{Calabrese:2007rg}
P.~Calabrese and J.~Cardy, ``{Quantum Quenches in Extended Systems},'' {\em J.
  Stat. Mech.}, vol.~0706, p.~P06008, 2007, 0704.1880.

\bibitem{Calabrese:2005in}
P.~Calabrese and J.~L. Cardy, ``{Evolution of entanglement entropy in
  one-dimensional systems},'' {\em J. Stat. Mech.}, vol.~0504, p.~P04010, 2005,
  cond-mat/0503393.

\bibitem{Cardy:2014rqa}
J.~Cardy, ``{Thermalization and Revivals after a Quantum Quench in Conformal
  Field Theory},'' {\em Phys. Rev. Lett.}, vol.~112, p.~220401, 2014,
  1403.3040.

\bibitem{AdS/CFT}
J.~M. Maldacena, ``{The Large N limit of superconformal field theories and
  supergravity},'' {\em Int. J. Theor. Phys.}, vol.~38, pp.~1113--1133, 1999,
  hep-th/9711200.

\bibitem{Gubser:1998bc}
S.~Gubser, I.~R. Klebanov, and A.~M. Polyakov, ``{Gauge theory correlators from
  noncritical string theory},'' {\em Phys. Lett. B}, vol.~428, pp.~105--114,
  1998, hep-th/9802109.

\bibitem{Witten:1998qj}
E.~Witten, ``{Anti-de Sitter space and holography},'' {\em Adv. Theor. Math.
  Phys.}, vol.~2, pp.~253--291, 1998, hep-th/9802150.

\bibitem{Ryu:2006bv}
S.~Ryu and T.~Takayanagi, ``{Holographic derivation of entanglement entropy
  from AdS/CFT},'' {\em Phys. Rev. Lett.}, vol.~96, p.~181602, 2006,
  hep-th/0603001.

\bibitem{Cardy:1988tk}
J.~L. Cardy and I.~Peschel, ``{Finite Size Dependence of the Free Energy in
  Two-dimensional Critical Systems},'' {\em Nucl. Phys. B}, vol.~300,
  pp.~377--392, 1988.

\bibitem{Holzhey:1994we}
C.~Holzhey, F.~Larsen, and F.~Wilczek, ``{Geometric and renormalized entropy in
  conformal field theory},'' {\em Nucl. Phys. B}, vol.~424, pp.~443--467, 1994,
  hep-th/9403108.

\bibitem{Korepin:2004zz}
V.~E. Korepin, ``Universality of entropy scaling in one dimensional gapless
  models,'' {\em Phys. Rev. Lett.}, vol.~92, p.~096402, Mar 2004.

\bibitem{CC}
P.~Calabrese and J.~L. Cardy, ``{Entanglement entropy and quantum field
  theory},'' {\em J. Stat. Mech.}, vol.~0406, p.~P06002, 2004, hep-th/0405152.

\bibitem{Danielsson:1999fa}
U.~H. Danielsson, E.~Keski-Vakkuri, and M.~Kruczenski, ``{Black hole formation
  in AdS and thermalization on the boundary},'' {\em JHEP}, vol.~02, p.~039,
  2000, hep-th/9912209.

\bibitem{Janik:2006gp}
R.~A. Janik and R.~B. Peschanski, ``{Gauge/gravity duality and thermalization
  of a boost-invariant perfect fluid},'' {\em Phys. Rev. D}, vol.~74,
  p.~046007, 2006, hep-th/0606149.

\bibitem{AbajoArrastia:2010yt}
J.~Abajo-Arrastia, J.~Aparicio, and E.~Lopez, ``{Holographic Evolution of
  Entanglement Entropy},'' {\em JHEP}, vol.~11, p.~149, 2010, 1006.4090.

\bibitem{Albash:2010mv}
T.~Albash and C.~V. Johnson, ``{Evolution of Holographic Entanglement Entropy
  after Thermal and Electromagnetic Quenches},'' {\em New J. Phys.}, vol.~13,
  p.~045017, 2011, 1008.3027.

\bibitem{Balasubramanian:2010ce}
V.~Balasubramanian, A.~Bernamonti, J.~de~Boer, N.~Copland, B.~Craps,
  E.~Keski-Vakkuri, B.~Muller, A.~Schafer, M.~Shigemori, and W.~Staessens,
  ``{Thermalization of Strongly Coupled Field Theories},'' {\em Phys. Rev.
  Lett.}, vol.~106, p.~191601, 2011, 1012.4753.

\bibitem{Aparicio:2011zy}
J.~Aparicio and E.~Lopez, ``{Evolution of Two-Point Functions from
  Holography},'' {\em JHEP}, vol.~12, p.~082, 2011, 1109.3571.

\bibitem{Keranen:2011xs}
V.~Keranen, E.~Keski-Vakkuri, and L.~Thorlacius, ``{Thermalization and
  entanglement following a non-relativistic holographic quench},'' {\em Phys.
  Rev. D}, vol.~85, p.~026005, 2012, 1110.5035.

\bibitem{Allais:2011ys}
A.~Allais and E.~Tonni, ``{Holographic evolution of the mutual information},''
  {\em JHEP}, vol.~01, p.~102, 2012, 1110.1607.

\bibitem{Buchel:2012gw}
A.~Buchel, L.~Lehner, and R.~C. Myers, ``{Thermal quenches in N=2* plasmas},''
  {\em JHEP}, vol.~08, p.~049, 2012, 1206.6785.

\bibitem{Buchel:2013gba}
A.~Buchel, R.~C. Myers, and A.~van Niekerk, ``{Universality of Abrupt
  Holographic Quenches},'' {\em Phys. Rev. Lett.}, vol.~111, p.~201602, 2013,
  1307.4740.

\bibitem{Hartman:2013qma}
T.~Hartman and J.~Maldacena, ``{Time Evolution of Entanglement Entropy from
  Black Hole Interiors},'' {\em JHEP}, vol.~05, p.~014, 2013, 1303.1080.

\bibitem{Liu:2013qca}
H.~Liu and S.~J. Suh, ``{Entanglement growth during thermalization in
  holographic systems},'' {\em Phys. Rev. D}, vol.~89, no.~6, p.~066012, 2014,
  1311.1200.

\bibitem{Bhaseen:2013ypa}
M.~Bhaseen, B.~Doyon, A.~Lucas, and K.~Schalm, ``{Far from equilibrium energy
  flow in quantum critical systems},'' {\em Nature Phys.}, vol.~11, p.~5, 2015,
  1311.3655.

\bibitem{Abajo-Arrastia:2014fma}
J.~Abajo-Arrastia, E.~da~Silva, E.~Lopez, J.~Mas, and A.~Serantes,
  ``{Holographic Relaxation of Finite Size Isolated Quantum Systems},'' {\em
  JHEP}, vol.~05, p.~126, 2014, 1403.2632.

\bibitem{daSilva:2014zva}
E.~da~Silva, E.~Lopez, J.~Mas, and A.~Serantes, ``{Collapse and Revival in
  Holographic Quenches},'' {\em JHEP}, vol.~04, p.~038, 2015, 1412.6002.

\bibitem{Caputa:2013eka}
P.~Caputa, G.~Mandal, and R.~Sinha, ``{Dynamical entanglement entropy with
  angular momentum and U(1) charge},'' {\em JHEP}, vol.~11, p.~052, 2013,
  1306.4974.

\bibitem{Takayanagi:2011zk}
T.~Takayanagi, ``{Holographic Dual of BCFT},'' {\em Phys. Rev. Lett.},
  vol.~107, p.~101602, 2011, 1105.5165.

\bibitem{Cardy:1989ir}
J.~L. Cardy, ``{Boundary Conditions, Fusion Rules and the Verlinde Formula},''
  {\em Nucl. Phys. B}, vol.~324, pp.~581--596, 1989.

\bibitem{Cardy:2004hm}
J.~L. Cardy, ``{Boundary conformal field theory},'' 11 2004, hep-th/0411189.

\bibitem{AdS/BCFT2}
M.~Fujita, T.~Takayanagi, and E.~Tonni, ``{Aspects of AdS/BCFT},'' {\em JHEP},
  vol.~11, p.~043, 2011, 1108.5152.

\bibitem{Cavalcanti:2018pta}
A.~G. Cavalcanti, D.~Melnikov, and M.~R. Silva, ``{Studies of Boundary Entropy
  in AdS/BCFT},'' {\em Class. Quant. Grav.}, vol.~37, no.~10, p.~105009, 2020,
  1808.07966.

\bibitem{Ugajin:2013xxa}
T.~Ugajin, ``{Two dimensional quantum quenches and holography},'' 11 2013,
  1311.2562.

\bibitem{Nozaki:2013wia}
M.~Nozaki, T.~Numasawa, and T.~Takayanagi, ``{Holographic Local Quenches and
  Entanglement Density},'' {\em JHEP}, vol.~05, p.~080, 2013, 1302.5703.

\bibitem{Mandal:2016cdw}
G.~Mandal, R.~Sinha, and T.~Ugajin, ``{Finite size effect on dynamical
  entanglement entropy: CFT and holography},'' 4 2016, 1604.07830.

\bibitem{Shimaji:2018czt}
T.~Shimaji, T.~Takayanagi, and Z.~Wei, ``{Holographic Quantum Circuits from
  Splitting/Joining Local Quenches},'' {\em JHEP}, vol.~03, p.~165, 2019,
  1812.01176.

\bibitem{Ageev:2019fjf}
D.~S. Ageev, ``{On the entanglement and complexity contours of excited states
  in the holographic CFT},'' 5 2019, 1905.06920.

\bibitem{Kudler-Flam:2020url}
J.~Kudler-Flam, Y.~Kusuki, and S.~Ryu, ``{Correlation measures and the
  entanglement wedge cross-section after quantum quenches in two-dimensional
  conformal field theories},'' {\em JHEP}, vol.~04, p.~074, 2020, 2001.05501.

\bibitem{Calabrese:2007mtj}
P.~Calabrese and J.~Cardy, ``{Entanglement and correlation functions following
  a local quench: a conformal field theory approach},'' {\em J. Stat. Mech.},
  vol.~0710, no.~10, p.~P10004, 2007, 0708.3750.

\bibitem{2008JSMTE..01..023E}
V.~{Eisler}, D.~{Karevski}, T.~{Platini}, and I.~{Peschel}, ``{Entanglement
  evolution after connecting finite to infinite quantum chains},'' {\em Journal
  of Statistical Mechanics: Theory and Experiment}, vol.~2008, p.~01023, Jan.
  2008, 0711.0289.

\bibitem{Stephan:2011}
J.-M. St\'ephan and J.~Dubail, ``Local quantum quenches in critical
  one-dimensional systems: entanglement, the loschmidt echo, and light-cone
  effects,'' {\em Journal of Statistical Mechanics: Theory and Experiment},
  vol.~2011, p.~P08019, Aug 2011, 1105.4846.

\bibitem{Stephan:2013}
J.-M. St\'ephan and J.~Dubail, ``Logarithmic corrections to the free energy
  from sharp corners with angle 2$\pi$,'' {\em Journal of Statistical
  Mechanics: Theory and Experiment}, vol.~2013, p.~P09002, Sep 2013, 1303.3633.

\bibitem{Asplund:2013zba}
C.~T. Asplund and A.~Bernamonti, ``{Mutual information after a local quench in
  conformal field theory},'' {\em Phys. Rev. D}, vol.~89, no.~6, p.~066015,
  2014, 1311.4173.

\bibitem{Asplund:2014coa}
C.~T. Asplund, A.~Bernamonti, F.~Galli, and T.~Hartman, ``{Holographic
  Entanglement Entropy from 2d CFT: Heavy States and Local Quenches},'' {\em
  JHEP}, vol.~02, p.~171, 2015, 1410.1392.

\bibitem{Karch:2000gx}
A.~Karch and L.~Randall, ``{Open and closed string interpretation of SUSY CFT's
  on branes with boundaries},'' {\em JHEP}, vol.~06, p.~063, 2001,
  hep-th/0105132.

\bibitem{DeWolfe:2001pq}
O.~DeWolfe, D.~Z. Freedman, and H.~Ooguri, ``{Holography and defect conformal
  field theories},'' {\em Phys. Rev. D}, vol.~66, p.~025009, 2002,
  hep-th/0111135.

\bibitem{Bak:2003jk}
D.~Bak, M.~Gutperle, and S.~Hirano, ``{A Dilatonic deformation of AdS(5) and
  its field theory dual},'' {\em JHEP}, vol.~05, p.~072, 2003, hep-th/0304129.

\bibitem{DHoker:2007zhm}
E.~D'Hoker, J.~Estes, and M.~Gutperle, ``{Exact half-BPS Type IIB interface
  solutions. I. Local solution and supersymmetric Janus},'' {\em JHEP},
  vol.~06, p.~021, 2007, 0705.0022.

\bibitem{DHoker:2007hhe}
E.~D'Hoker, J.~Estes, and M.~Gutperle, ``{Exact half-BPS Type IIB interface
  solutions. II. Flux solutions and multi-Janus},'' {\em JHEP}, vol.~06,
  p.~022, 2007, 0705.0024.

\bibitem{Aharony:2011yc}
O.~Aharony, L.~Berdichevsky, M.~Berkooz, and I.~Shamir, ``{Near-horizon
  solutions for D3-branes ending on 5-branes},'' {\em Phys. Rev. D}, vol.~84,
  p.~126003, 2011, 1106.1870.

\bibitem{Magan:2014dwa}
J.~M. Mag\'an, D.~Melnikov, and M.~R.~O. Silva, ``{Black Holes in AdS/BCFT and
  Fluid/Gravity Correspondence},'' {\em JHEP}, vol.~11, p.~069, 2014,
  1408.2580.

\bibitem{Nozaki_2012}
M.~Nozaki, T.~Takayanagi, and T.~Ugajin, ``Central charges for bcfts and
  holography,'' {\em Journal of High Energy Physics}, vol.~2012, Jun 2012.

\bibitem{Erdmenger_2015}
J.~Erdmenger, M.~Flory, and M.-N. Newrzella, ``Bending branes for dcft in two
  dimensions,'' {\em Journal of High Energy Physics}, vol.~2015, Jan 2015.

\bibitem{Seminara:2017hhh}
D.~Seminara, J.~Sisti, and E.~Tonni, ``{Corner contributions to holographic
  entanglement entropy in AdS$_{4}$/BCFT$_{3}$},'' {\em JHEP}, vol.~11, p.~076,
  2017, 1708.05080.

\bibitem{Seminara:2018pmr}
D.~Seminara, J.~Sisti, and E.~Tonni, ``{Holographic entanglement entropy in
  AdS$_{4}$/BCFT$_{3}$ and the Willmore functional},'' {\em JHEP}, vol.~08,
  p.~164, 2018, 1805.11551.

\bibitem{Shashi:2020mkd}
S.~Shashi, ``{Quotient-AdS/BCFT: Holographic Boundary CFT$_2$ on AdS$_3$
  Quotients},'' 5 2020, 2005.10244.

\bibitem{Sato:2020upl}
Y.~Sato, ``{Boundary entropy under ambient RG flow in the AdS/BCFT model},''
  {\em Phys. Rev. D}, vol.~101, no.~12, p.~126004, 2020, 2004.04929.

\bibitem{Berenstein:1998ij}
D.~E. Berenstein, R.~Corrado, W.~Fischler, and J.~M. Maldacena, ``{The Operator
  product expansion for Wilson loops and surfaces in the large N limit},'' {\em
  Phys. Rev. D}, vol.~59, p.~105023, 1999, hep-th/9809188.

\bibitem{Caputa:2019avh}
P.~Caputa, T.~Numasawa, T.~Shimaji, T.~Takayanagi, and Z.~Wei, ``{Double Local
  Quenches in 2D CFTs and Gravitational Force},'' {\em JHEP}, vol.~09, p.~018,
  2019, 1905.08265.

\bibitem{Akal:2020wfl}
I.~Akal, Y.~Kusuki, T.~Takayanagi, and Z.~Wei, ``{Codimension two holography
  for wedges},'' 7 2020, 2007.06800.

\bibitem{Hubeny:2007xt}
V.~E. Hubeny, M.~Rangamani, and T.~Takayanagi, ``{A Covariant holographic
  entanglement entropy proposal},'' {\em JHEP}, vol.~07, p.~062, 2007,
  0705.0016.

\bibitem{Erdmenger_2016entang}
J.~Erdmenger, M.~Flory, C.~Hoyos, M.-N. Newrzella, and J.~M.~S. Wu,
  ``Entanglement entropy in a holographic kondo model,'' {\em Fortschritte der
  Physik}, vol.~64, p.~109–130, Jan 2016.

\bibitem{Erdmenger_2016holog}
J.~Erdmenger, M.~Flory, C.~Hoyos, M.-N. Newrzella, A.~O’Bannon, and J.~M.~S.
  Wu, ``Holographic impurities and kondo effect,'' {\em Fortschritte der
  Physik}, vol.~64, p.~322–329, Mar 2016.

\bibitem{Miao_2017}
R.-X. Miao, C.-S. Chu, and W.-Z. Guo, ``New proposal for a holographic boundary
  conformal field theory,'' {\em Physical Review D}, vol.~96, Aug 2017.

\bibitem{Brown:1986nw}
J.~Brown and M.~Henneaux, ``{Central Charges in the Canonical Realization of
  Asymptotic Symmetries: An Example from Three-Dimensional Gravity},'' {\em
  Commun. Math. Phys.}, vol.~104, pp.~207--226, 1986.

\bibitem{Wen:2018svb}
X.~Wen, S.~Ryu, and A.~W. Ludwig, ``{Entanglement hamiltonian evolution during
  thermalization in conformal field theory},'' {\em J. Stat. Mech.}, vol.~1811,
  no.~11, p.~113103, 2018, 1807.04440.

\end{thebibliography}

\end{document}